\documentclass[nofootinbib,superscriptaddress]{revtex4}
\pdfoutput=1
\usepackage{amsmath,amssymb}
\usepackage{epsfig}
\usepackage{hyperref}
\usepackage{color}
\usepackage{slashed}
\usepackage{placeins}

\usepackage{amsfonts}
\usepackage{graphicx, rotating}
\usepackage{epstopdf}
\usepackage{latexsym}
\usepackage{rotating}

\usepackage[font={small}]{caption}   

\definecolor{myred}{rgb}{0.7, 0, 0}
\definecolor{myblue}{rgb}{0, 0, 0.7}
\definecolor{mygreen}{rgb}{0.04, 0.7, 0.5}
\definecolor{mygray}{rgb}{0.1, 0.1, 0.1}

\hypersetup{colorlinks,citecolor=myred,linkcolor=myblue,urlcolor=myblue,linktocpage=true}

 \def\be   {\begin{equation}}   \def\ee   {\end{equation}}
 \def\ba   {\begin{array}}      \def\ea   {\end{array}}
 \def\bea  {\begin{eqnarray}}   \def\eea  {\end{eqnarray}}
 \def\bean {\begin{eqnarray*}}  \def\eean {\end{eqnarray*}}
 \def\nn{\nonumber}
 \def\bry{\begin{array}}
 \def\ery{\end{array}}

\numberwithin{equation}{section}

\begin{document}

\vspace{-2cm}

\begin{flushright}
DESY 19-207
\end{flushright}
\title{
High-Temperature Electroweak Symmetry Non-Restoration \\
\vspace{.2cm}
from New Fermions and Implications for Baryogenesis
}
\author{Oleksii Matsedonskyi}
\address{Department of Particle Physics and Astrophysics, Weizmann Institute of Science, Rehovot 761001, Israel}
\author{G\'eraldine Servant}
\address{DESY, Notkestrasse 85, 22607 Hamburg, Germany \\
II.~Institute of Theoretical Physics, University of Hamburg, 22761 Hamburg, Germany}

%
\begin{abstract}

The strength of electroweak symmetry breaking may substantially differ in the early Universe compared to the present day value. 
In the Standard Model,  the Higgs vacuum expectation value ({\it vev}) vanishes and electroweak symmetry gets restored at temperatures above $\sim 160$ GeV due to the Higgs field interactions with the high-temperature plasma. It was however shown that new light singlet scalar fields may change this behaviour.
The key feature is the non-standard dependence on the Higgs {\it vev} of the new particles mass which 
can vanish at large Higgs {\it vev}, inducing a negative correction to the Higgs thermal mass, leading to electroweak symmetry non-restoration at high temperature. 
We show that such an effect can also be induced by new singlet fermions which on the other hand have the advantage of not producing unstable directions in the scalar potential at tree level, nor bringing additional severe hierarchy problems.  
As temperature drops, such a high-temperature breaking phase may continuously evolve into the zero-temperature breaking phase or the two phases can be separated by a temporary phase of restored symmetry.  We discuss how our construction can naturally arise in motivated models of new physics, such as Composite Higgs. This is particularly relevant for baryogenesis, as it opens a whole class of possibilities in which the baryon asymmetry can be produced  during a high temperature phase transition, while not being erased later by sphalerons.

\end{abstract}
%
%

\maketitle


\tableofcontents



\section{Introduction}

The Standard Model (SM) predicts electroweak (EW) symmetry restoration at high temperature due to the large positive thermal corrections to the Higgs mass parameter, coming from the Higgs boson interactions in the hot plasma, mainly  with the top quark, the electroweak gauge bosons and the Higgs boson itself.
It is however interesting to analyse the possibility of EW symmetry non-restoration (SNR)  for several reasons. In the following, we will be mostly interested in a specific type of SNR (so-called \emph {continuous} SNR) in which the EW symmetry remains broken from some high temperature (larger than $\sim$ 160 GeV) down to $T=0$. Moreover, the value of the Higgs field remains larger than the temperature, $h/T\gtrsim 1$. Such a specific type of SNR is very important for electroweak baryogenesis.   
In this framework, the baryon asymmetry can be generated during a first-order phase transition in the early universe, but only if this transition results in the growth of the Higgs vacuum expectation value ({\it vev}) to a value higher than the temperature. 
For instance, a number of UV completions of the SM contain new scalars above the EW scale, which can undergo such phase transitions, and are coupled to the Higgs field. However, if the critical temperature of the transition it too high, the electroweak symmetry would remain unbroken and no baryon asymmetry can be generated, unless the phase transition is supercooled. But in the latter case,
any produced baryon asymmetry will still be washed out by sphalerons after reheating in ($B-L$) conserving theories if the reheat temperature is too high.  
For this reason, EW baryogenesis is generally thought to be tied to happen at $T\sim {\cal O}(100)$ GeV, allowing for $h/T\gtrsim 1$. If, on the other hand, EW symmetry is broken by some new high-temperature effects, the baryon asymmetry produced during the high-scale phase transition can be preserved. This makes it possible to use new sources of CP-violation without conflict with experimental bounds on electric dipole moments~\cite{Andreev:2018ayy}, as well as heavier and less constrained sectors inducing the first-order EW phase transition. High-temperature EW SNR is therefore highly relevant, although it has so far only been scarcely addressed in the literature.
    
High-temperature SNR was first discussed by Weinberg in the simple two-scalar model~\cite{Weinberg:1974hy}, and by Mohapatra and Senjanovic in connection with non-restoration of CP symmetry \cite{Mohapatra:1979qt,Mohapatra:1979bt} (see Ref.~\cite{Fujimoto:1984hr,Salomonson:1984rh,Salomonson:1984px,Dvali:1995cj,Bimonte:1995xs,Bimonte:1995sc,Dvali:1996zr,Orloff:1996yn,Pietroni:1996zj,Gavela:1998ux,Bajc:1998jr,Jansen:1998rj,Bimonte:1999tw,Pinto:1999pg,Aziz:2009hk,Kilic:2015joa} for the subsequent works). But only recently the idea was applied to the EW symmetry, see Ref.~\cite{Meade:2018saz,Baldes:2018nel,Glioti:2018roy}. In these models, high-temperature EW SNR  is driven by a new sector containing a large number of relatively light singlet scalars interacting with the Higgs doublet.
In this work, we explore the phenomenon of EW SNR driven by new fermionic degrees of freedom. The general underlying principle for SNR driven by new particles is fairly simple. Massless, or sufficiently light ($m\lesssim T$) particles coupled to the Higgs produce a dip in the thermal Higgs effective potential of the size $\delta V \propto - T^4$. On the other hand, heavy particles ($m\gg T$) have a negligible contribution. Having this in mind, we will construct models which feature new singlet fermions
with a specific Higgs-dependent mass. This mass is  sizeable at zero Higgs {\it vev} $h=0$ and vanishes at some large $h$. We then find that the plasma containing such fermions induces a correction to the Higgs potential which is minimized at large $h$ and is able to trigger SNR~\footnote{While in this paper we will concentrate on the case where the only scalar field responsible for EWSB is the Higgs field, our construction can be straightforwardly extended to SNR due to non-zero {\it vev} of an additional scalar that is charged under the electroweak symmetry in the spirit of Ref.~\cite{Patel:2012pi,Blinov:2015sna,Inoue:2015pza}. Such a possibility was already discussed in the context of SNR with new scalars in Ref.~\cite{Glioti:2018roy}.}. Such a mass dependence is opposite to the one featured by the SM fermions, and new fermions will have to be introduced. As their effect on the Higgs field has to remain sizeable at $T\simeq160$~GeV, where the SM thermal effects would otherwise restore the EW symmetry, the zero-$T$ mass of such fermions has to be of the same order, {\it i.e.} at most a few hundreds of GeV. 

The fermionic- and the previously considered scalar-induced SNR have several important differences. First, having new light fermions is {\it a priori} less troublesome from the naturalness point of view. Second, the fermions do not alter the tree-level scalar potential of the model. The new scalars responsible for SNR do modify it in such a way that the potential becomes unstable at large $h$ values, unless a severe constraint is imposed on the number of scalars, which typically has to exceed a few hundreds.        
On the other hand, as we will see in Sec.~\ref{sec:snrwithferm}, fermionic SNR is generically linked to non-renormalizable operators, and therefore has an intrinsic energy and temperature cutoff, above which the SNR effect disappears or at least the model loses perturbativity. Even though the mass of new fermions will feature a linear sensitivity to the cutoff, regulating such a sensitivity is a much simpler task than ensuring a lightness of new scalar degrees of freedom.

To show this effect, we will use a simplified model which features only the minimal number of necessary  ingredients for SNR -- $n$ copies of a singlet fermion $N$ coupled to the Higgs through a dimension-five operator.  
This model can however easily be embedded into more appealing UV completions. We discuss two such completions -- the models of Goldstone Higgs and a singlet+doublet extension of the SM. As for the former case, high-$T$ SNR has already been discussed in the context of the Little Higgs models in Ref.~\cite{Espinosa:2004pn}. However, in that case, the SNR was supposed to happen only at some high $T\gg m_W$ with a questionable validity of one-loop predictions~\cite{Ahriche:2010kh}. An attempt to achieve a continuous SNR in CH models was reported in~\cite{DiLuzio:2019wsw}, with no viable parameter space found.   
The main difference of our CH construction, which allows for a perturbatively controlled SNR, 
is the presence of a large, at least ${\cal O}(10)$, number of new fermions with sufficiently unconstrained couplings.

We will start with a general discussion of the temperature corrections to the Higgs potential  in Sec.~\ref{sec:snr} and identify the main ingredients needed to induce SNR with fermions, pointing to a simplified model with singlet Dirac fermions containing a Higgs-dependent dimension-5 mass term. We dedicate Sec.~\ref{sec:paramspace} to a more refined analysis, including the estimate of the higher-loop effects, and a numerical computation of the Higgs {\it vev} temperature evolution. The UV completions to the simplified model are presented in Sec.~\ref{sec:explmod}. Sec.~\ref{sec:tracking} contains a detailed comparison between scalar and fermionic SNR.  We summarize our results in Sec.~\ref{sec:conc}. In appendices~\ref{eq:appsm} and~\ref{sec:deltamn} we detail the SM thermal corrections and the thermal corrections to the new fermions mass.

\section{Thermal Corrections and Symmetry Non-Restoration}\label{sec:snr}

\subsection{One-Loop Thermal Corrections}

The Standard Model Higgs doublet induces spontaneous breaking of the EW symmetry at zero temperature, provided by a negative mass\footnote{Mass squared should be understood whenever we mention negative scalar mass.} parameter in the scalar potential  
\be\label{eq:vsmt0}
V_h^{\text{SM}} = -\frac{\mu^2} 2 h^2 + \frac{\lambda}{4} h^4,
\ee
where $h$ denotes the average value of the Higgs field, $\mu \simeq 90$~GeV and $\lambda \simeq 0.13$, with $ \langle h \rangle =v_{\rm SM}=246$~GeV and $m_h^2 = 126$~GeV at the $V_h^{\text{SM}}$ minimum.  The effect of the Higgs field interaction with high-temperature plasma can be accounted for by the higher-order corrections to the Higgs potential.
The leading ``one-loop'' thermal corrections are given by
\be\label{eq:1loopvh}
\Delta V_b^T = \frac{T^4}{2\pi^2} J_b[{m^2}/{T^2}],\qquad
\Delta V_f^T = -\frac{2T^4}{\pi^2} J_f[{m^2}/{T^2}]
\ee
respectively for one thermalized bosonic degree of freedom and one Dirac fermion with mass $m$. Their interactions with the Higgs field are encoded in the $h$-dependent masses $m=m(h)$. The thermal loop functions are defined as
\be
J_b[x]= \int_0^\infty dk\, k^2 \log \left[1- e^{-\sqrt{k^2+x}} \right],\qquad
J_f[x]= \int_0^\infty dk\, k^2 \log \left[1+ e^{-\sqrt{k^2+x}} \right].
\ee 
The corrections~(\ref{eq:1loopvh}) have minima at $m^2=0$ (within $m^2\geq 0$ region). In the high-temperature limit $m^2/T^2\ll 1$ they simplify to
\be\label{eq:Vbfexp}
\Delta V_b^T \simeq -\frac{\pi^2 T^4}{90}  + \frac{T^2 m^2}{24} ,\qquad
\Delta V_f^T \simeq -\frac{7\pi^2 T^4}{180}  + \frac{T^2 m^2}{12}.
\ee
The first terms of the expansions~(\ref{eq:Vbfexp}) define the depth of the negative correction to the Higgs potential at $m^2=0$. The second terms set the size of the correction to Higgs mass in the vicinity of the minimum, which is given by

\be
\boxed{\delta m_h^2(T) \,\propto\, \left. T^2 (m^2(h))^{\prime \prime}\right|_{m=0}}
\ee 

\begin{figure}[t]
\includegraphics[width=8.cm]{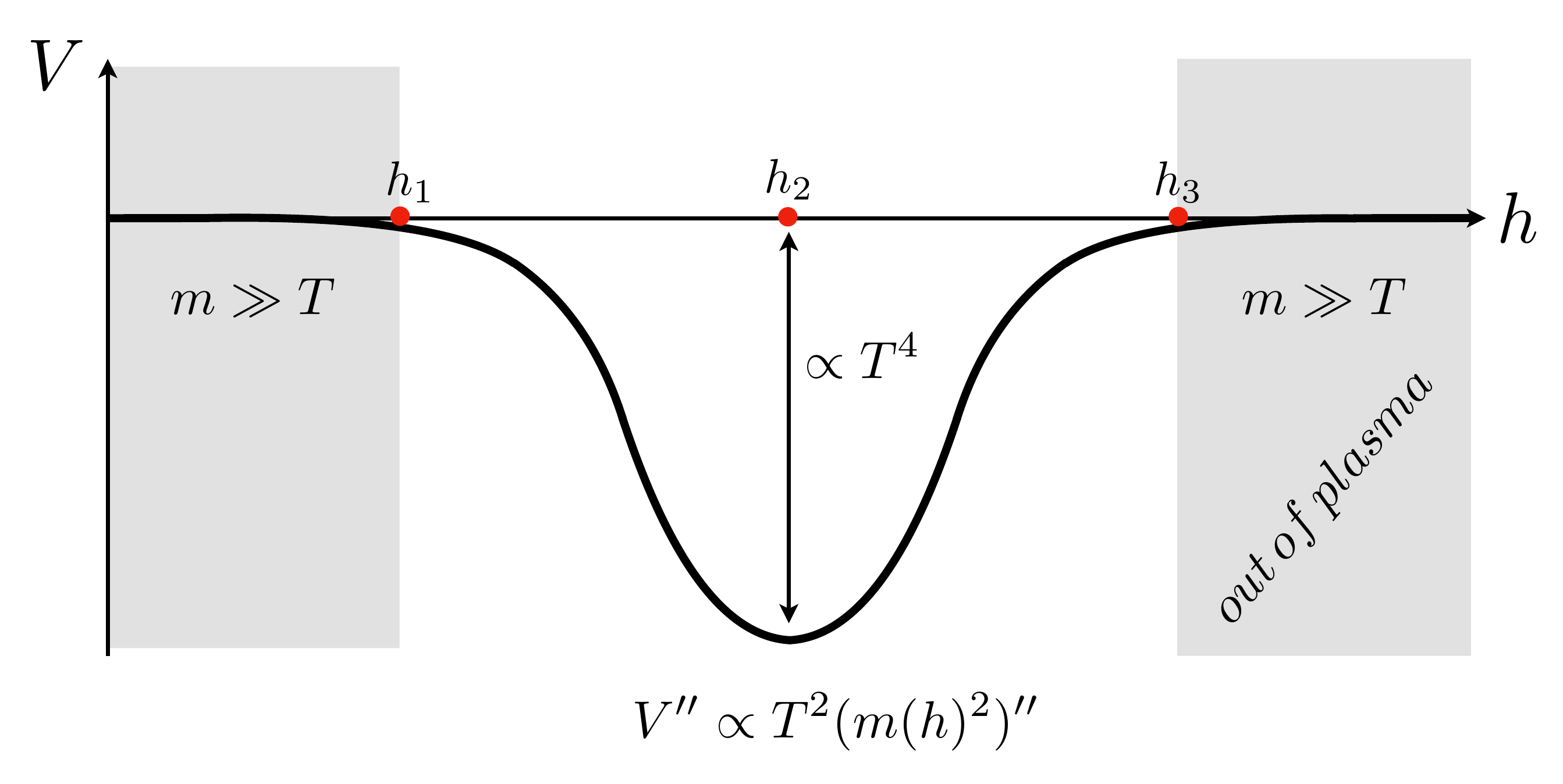}
\hspace{0.5cm}
\includegraphics[width=8.cm]{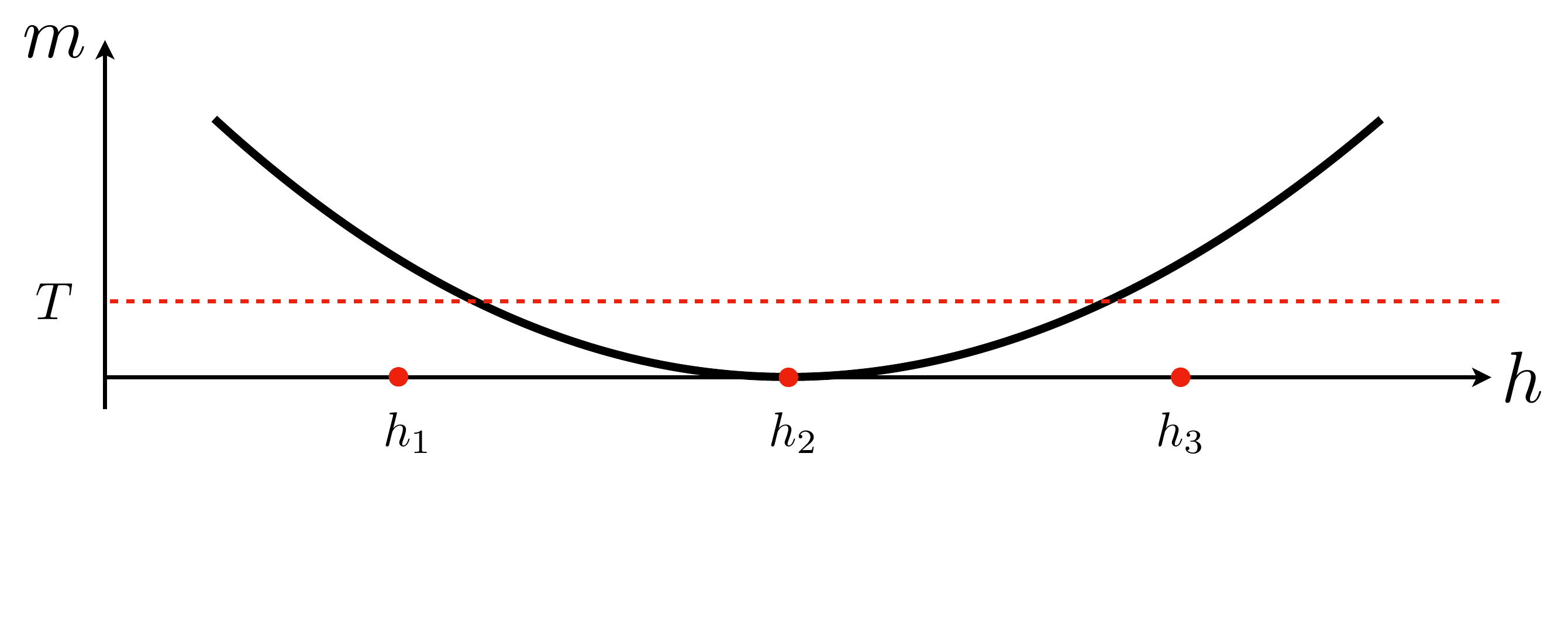}
\caption{Sketch of the thermal correction to the Higgs potential (left panel) induced by particles in the plasma whose mass depends on the Higgs field as shown on the right panel. At high temperature, a dip is induced in the potential at the point where the particle mass term vanishes.}
\label{fig:scheme}
\end{figure}

On the other hand, for $m^2/T^2 \gg 1$ the thermal corrections vanish. The corresponding schematic picture of the one-loop thermal correction is shown in Fig.~\ref{fig:scheme}. 
In that figure, we assume that the particle mass  gradually decreases with $h$, reaches zero and then increases. Such a behaviour is easy to realize for fermionic mass terms, which we concentrate on in this work. The plots in Fig.~\ref{fig:scheme} are only partly applicable to the case of scalar fields, as their squared mass would typically become negative after reaching zero, leading to an instability.
Now, the two most important aspects to analyse are where the $m(h)=0$ point is located and how steep are the walls around the dip, in other words, what is the size of the induced correction to the Higgs mass around the dip. 

In the Standard Model, the masses of the elementary states vanish at $h=0$. Corresponding thermal corrections have a dip around $h=0$, which grows with temperature and eventually becomes a global minimum of the potential.  The resulting thermal correction to the Higgs mass in the $m^2/T^2 \ll 1$ limit is given by
\be\label{eq:vhexpsm}
\delta m^2_h(T) \simeq  T^2 \left[
\frac{\lambda_t^2}{4} +\frac{\lambda}{2} + \frac{3g^2}{16} + \frac{g^{\prime 2}}{16} 
\right] \simeq  0.4 \ T^2 ,
\ee
where $\lambda_t$ is the top quark Yukawa coupling and $g, g^\prime$ are EW gauge couplings. 
This correction is dominated by the contribution of the top quark. In this picture, the relative strength of EW symmetry breaking, $h/T$, drops below 1 at $T\gtrsim 130$~GeV and EW symmetry gets restored at $T\gtrsim 160$~GeV~\cite{DOnofrio:2014rug}. 

We therefore want to investigate which type of modifications to the SM does not lead to this symmetry restoration, using the thermal effects of fermionic fields.

\subsection{Modified Standard Model Interactions}
\label{sec:modsm}

 Following the path of gradual increase of complexity, we start by considering the case of the SM effective field theory, i.e. the theory featuring the SM states only, but containing higher-dimensional operators. One of the simplest ways to change the picture described above is for instance to modify the SM Yukawa interactions to make the fermion mass vanish at some large Higgs {\it vev}, e.g. 
\be\label{eq:qqhhh}
{\cal L}_{\text{Yuk}} = - \lambda_q \bar q h q (1 - h^2/f^2).
\ee 
where $\lambda_q$ is the Yukawa coupling and $f$ is some mass scale suppressing the dimension-six operator.
In such a case, the contribution of the $q$ quark to the Higgs thermal potential would have two minima, at the points where $m_q=0$: one at $h=0$ and another at $h \sim f$, suggesting a possibility of symmetry non-restoration. The first subtlety here is that for $h\sim f$ the effective field theory expansion in the powers of $h/f$ breaks down. To make any predictions in this regime one needs to invoke some type of UV completion for Eq.~(\ref{eq:qqhhh}). One simple example would be the models with a Higgs being a pseudo Nambu-Goldstone boson (PNGB), arising e.g. as a pion-like state of some new strongly interacting sector. We discuss this option in detail in Sec.~\ref{sec:nghiggs}. PNGBs can be conveniently parametrized as phases of trigonometric functions and the term responsible for the quark mass can for instance take the form 
\be
m_q \sim \lambda_q f \sin (h/f) \cos (h/f).
\ee 
The absolute value of the mass (we are not interested in the phase of the fermionic mass terms, as it can be rotated away) has two minima, at $h=0$ and $h=\pi f/2$. One should however keep in mind that both minima are of the same depth 
\be
\Delta V_f^T \simeq -\frac{7\pi^2 T^4}{180},
\ee
see Eq.~(\ref{eq:Vbfexp}). Other thermal corrections (e.g. from the SM gauge bosons) and the zero-temperature potential typically make the $h=0$ minimum deeper. Therefore SNR is not expected to occur, and we have to consider adding new fermions instead of simply modifying the SM couplings. Nevertheless, the effect of modified Yukawas is important, as it can facilitate SNR by reducing the SM contribution (e.g. the large correction from the top quark) to the thermal potential at large $h$. Moreover, such Yukawa modifications are automatically present in some beyond-the-Standard-Model constructions, as we will see in Sec.~\ref{sec:nghiggs}. We should therefore keep in mind that they play a relevant role.

\subsection{Symmetry Non-Restoration with New Fermions}
\label{sec:snrwithferm}

\begin{figure}[t]
\includegraphics[width=8.cm]{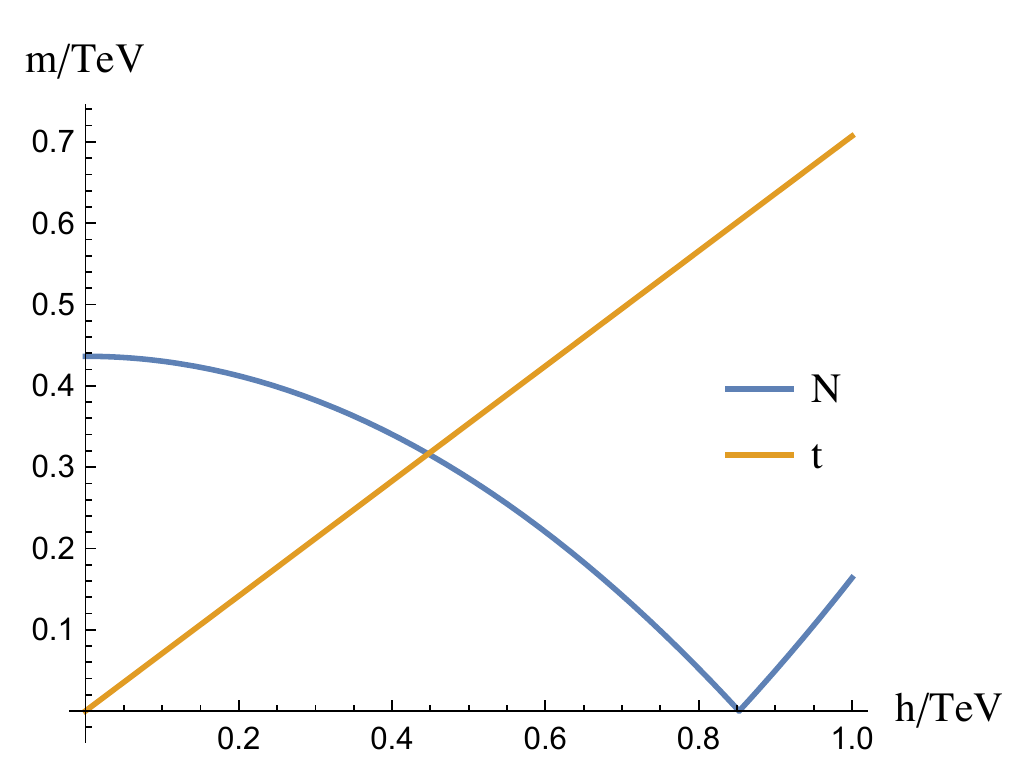}
\hspace{0.5cm}
\includegraphics[width=8.cm]{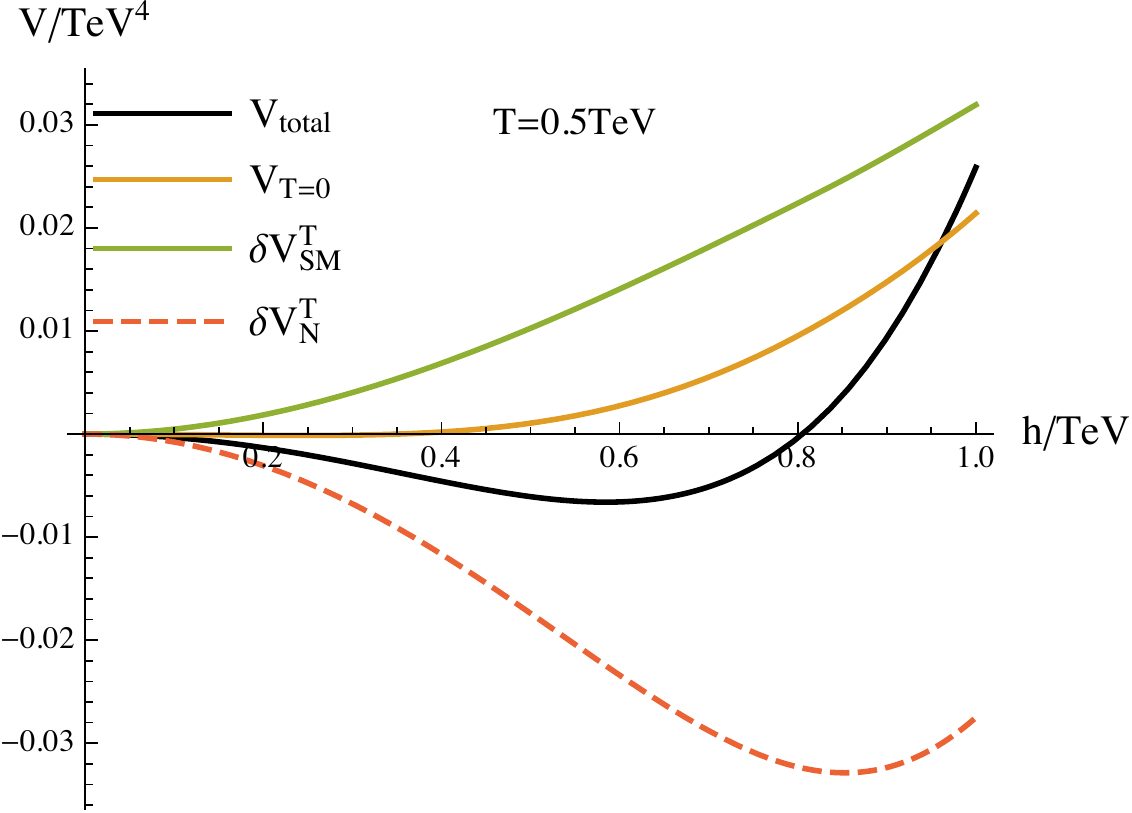}
\caption{{\bf Left}: Top quark mass (orange) and the $N$ fermion mass, which is minimized at large Higgs {\it vev} (blue). {\bf Right}: Corresponding 1-loop Higgs thermal potential featuring SNR at $T=0.5$~TeV (black solid) and its decomposition into non-thermal part (orange solid), finite temperature corrections from the SM interactions (green solid) and from the interactions with the $N$ fermions (red dashed). The maximal negative correction from the $N$ fermions is at the point of vanishing $N$ mass corresponding to large Higgs {\it vev}. 
For these plots we chose $n=10$, $\Lambda=1$~TeV, $\lambda_N=0.6$, $m_N(v_{\rm SM})=0.4$~TeV.}
\label{fig:example}
\end{figure}

Let us now add new fermions with a Higgs-dependent mass to the model. The simplest case is a singlet Dirac fermion $N$ coming in $n$ copies. The Lagrangian leading to SNR is 
\be\label{eq:LN}
\boxed{{\cal L}_N = -m_N^{(0)} \bar N N + \lambda_N  \bar N N h^2/\Lambda }
\ee  
where $\Lambda$ is the scale at which our effective field theory (EFT) is UV-completed by some heavier states, $\lambda_N$ is a positive coupling and $m_N$ is a positive mass parameter.
The dip in the thermal correction to the Higgs potential appears at the point of vanishing $N$ mass (see Fig.~\ref{fig:example})
\be\label{eq:mN0}
m_N(h) = m_N^{(0)} - \lambda_N h^2/\Lambda = 0\quad \longrightarrow \quad h^2 = m_N^{(0)} \Lambda/\lambda_N.
\ee
Around the Higgs field origin, the negative correction to the Higgs mass in the $m_N\ll T$ limit is approximately given by 
\be\label{eq:deltamh0}
\delta m_h^2 [T] \simeq 
n \frac {T^2}{12} (m_N^2(h))^{\prime \prime} = 
   -  n \lambda_N  \frac{m_N^{(0)}} {3 \Lambda} T^2.
\ee
This negative correction to the Higgs mass, if large enough, can surpass the positive SM thermal corrections and eventually make the Higgs field origin unstable, leading to high temperature SNR. Comparing Eq.s~(\ref{eq:vhexpsm}) and~(\ref{eq:deltamh0}), we find the necessary condition for this to happen
\be\label{eq:nNcond1}
n \lambda_{N}  \gtrsim 5 \left(\frac {v_{\text{SM}}} {m_N^{(0)}} \right) \left(\frac{\Lambda} {\text{TeV}}  \right) \ \ \mbox{or, equivalently,} \ \ \boxed{n \lambda_{N} \frac{m_N^{(0)}}{\Lambda} \gtrsim 1 .}
\ee
This SNR condition is only valid when the new fermions contribute significantly to the plasma density, i.e. 
\be
\boxed{m_N (h\simeq0) \lesssim T.}
\ee 
Otherwise the $N$-induced correction is suppressed. For this reason, having SNR not only at some high temperature, but also at the temperatures around the EW scale, requires $N$ to be relatively light. On the other hand, the fermion mass is also the parameter which enhances the negative Higgs mass correction (\ref{eq:deltamh0}), and therefore it cannot be too small either. 
Fig.~\ref{fig:example} shows, for some choice of parameters, how the addition of weak-scale fermions induces EW SNR behaviour at high temperature.
The components of the plotted potential
\be
V_{total} = V_{T=0} + \delta V_{\text{SM}}^T + \delta V_{N}^T
\ee 
are discussed in the next section. The zero-temperature potential $V_{T=0}$ consists of the tree-level potential~(\ref{eq:vsmt0}) and one-loop corrections induced by the SM states~(\ref{eq:v1lsm}) and by the new fermions~(\ref{eq:deltav1loop}). The SM thermal correction $\delta V_{\text{SM}}^T$ is given in Eq.~(\ref{eq:vhtfullsm}).  Inclusion of the $T=0$ loop correction (which decreases the Higgs quartic) and the full thermal correction from the SM states (which tends to become flat at $h\gg T$, contrary to the leading quadratic piece in Eq.~(\ref{eq:vhexpsm})), both facilitate shifting the minimum closer to large $h$. 
The thermal correction from the $N$ fermions $\delta V_{N}^T$ is given in Eq.~(\ref{eq:1loopvh}) and is the dominant effect.

\begin{figure}[t]
\includegraphics[width=0.24\textwidth]{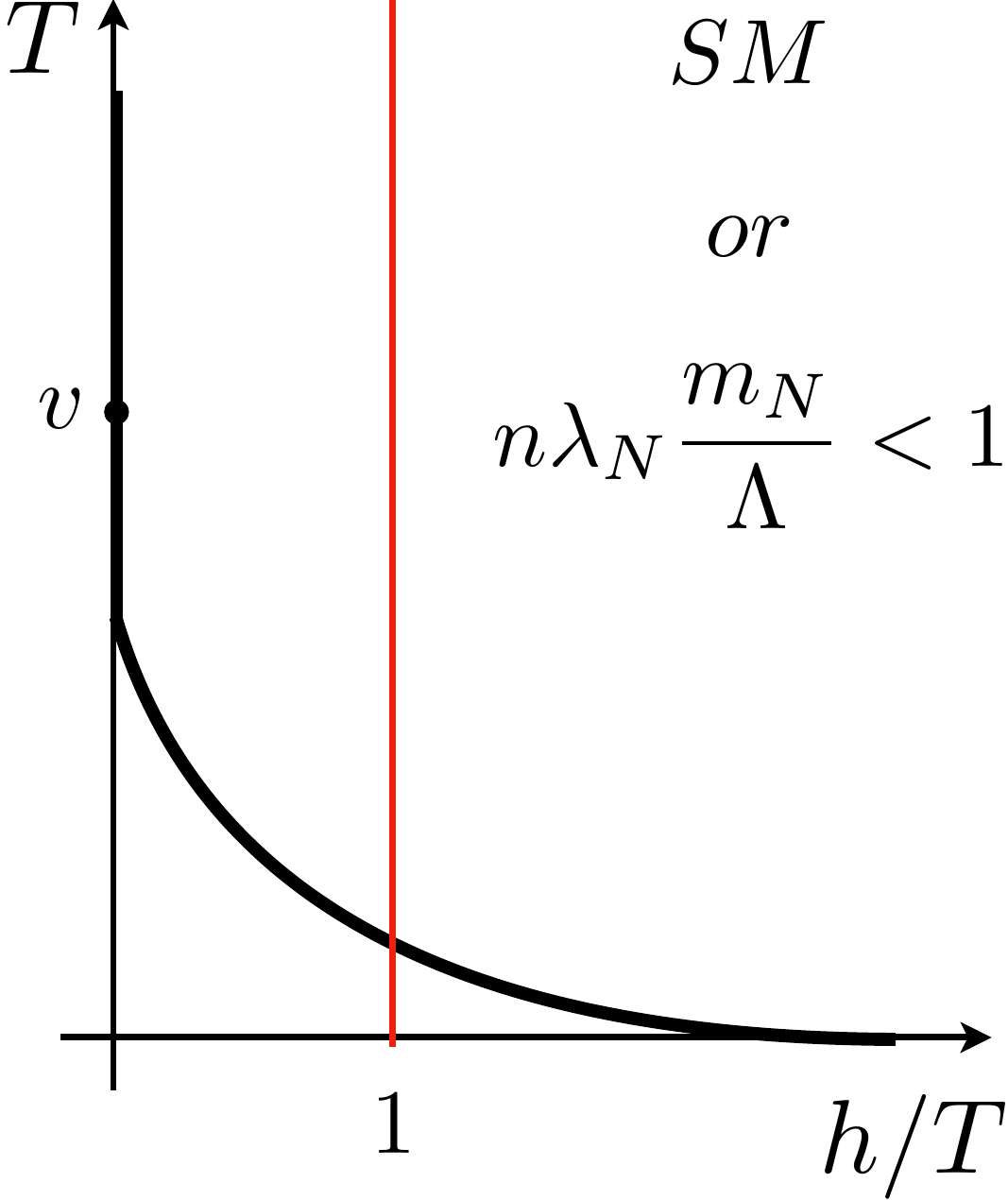}
\hspace{1.5cm}
\includegraphics[width=0.255\textwidth]{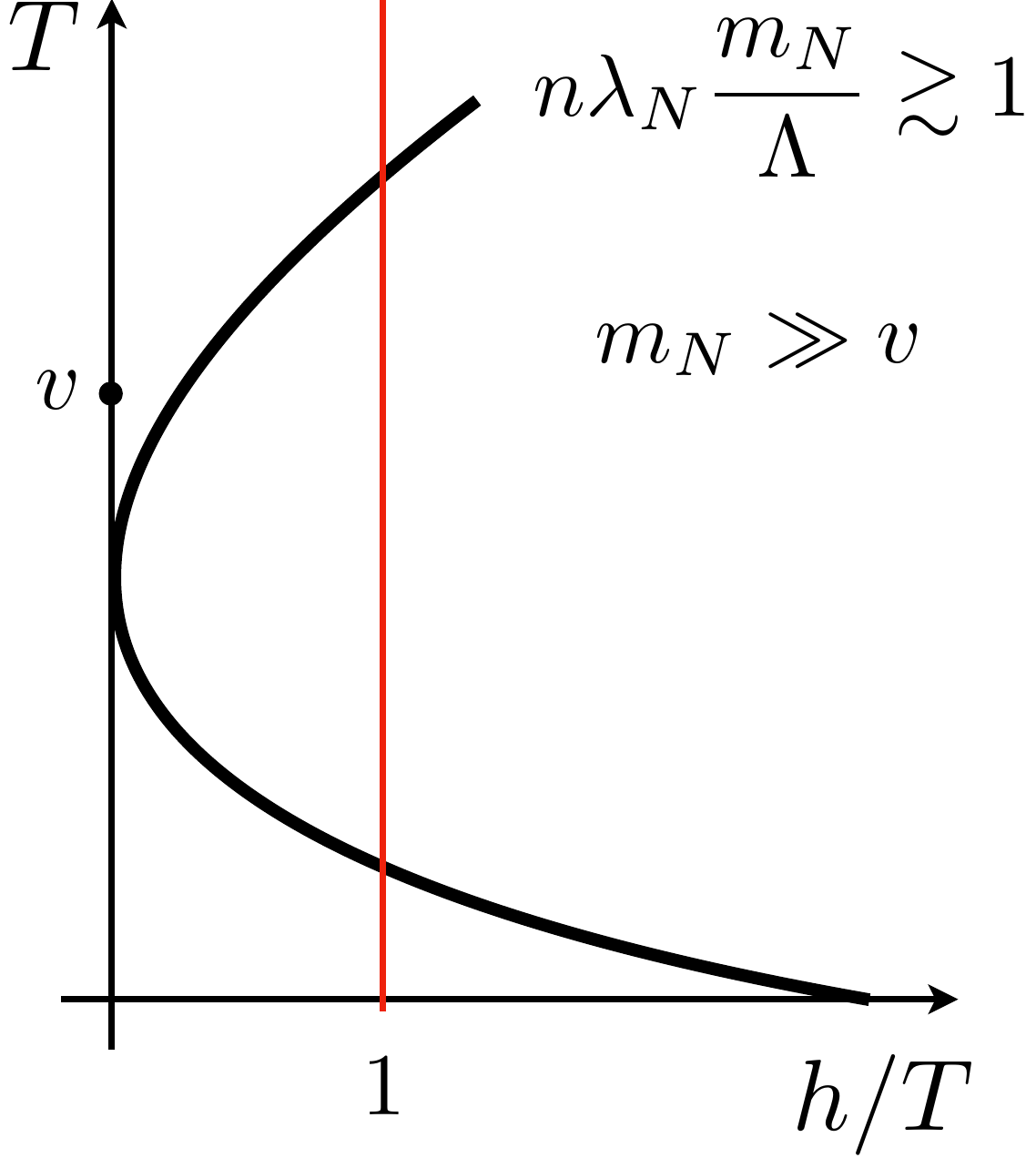}
\hspace{1.5cm}
\includegraphics[width=0.25\textwidth]{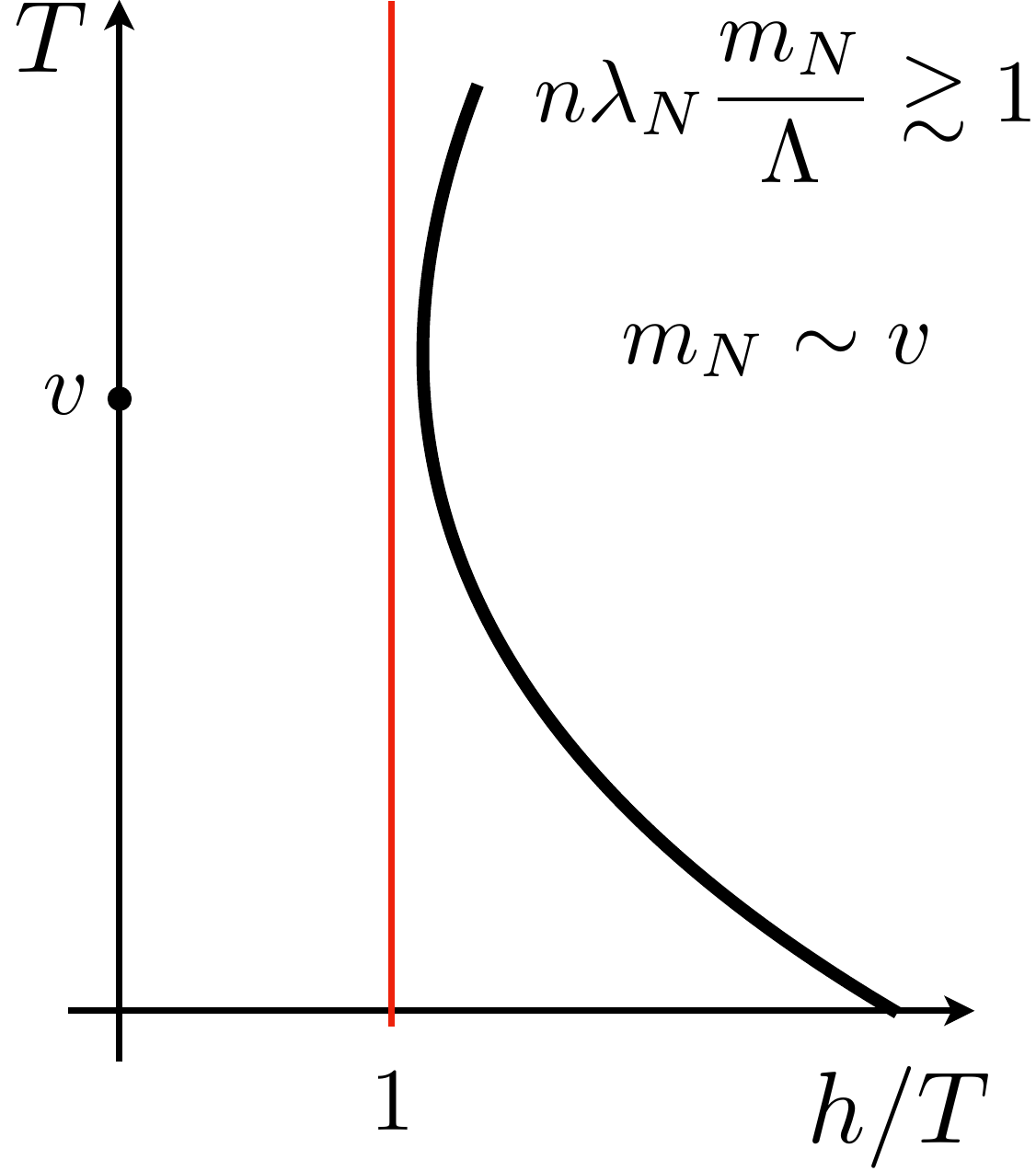}
\caption{Schematic plots of $h/T$ dependence on the temperature.
{\bf Left}:  Behaviour found in the SM, or in a model with new fermions where the SNR condition~(\ref{eq:nNcond1}) is not met.
{\bf Center}: Model with new fermions where the SNR condition is satisfied, but the fermions are too heavy to affect the Higgs potential at temperatures around the EW scale.
{\bf Right}: Model with new fermions satisfying the SNR condition and light enough to contribute to the Higgs potential at temperatures around the EW scale, such that the sphaleron bound $h/T\gtrsim 1$ is always satisfied. 
For both the center and right plots, we have assumed that the position of the minimum of the thermal part of the potential induced by the new fermions, $h^2 = m_N^{(0)} \Lambda/\lambda_N$, is always greater than $T^2$ within the plotted temperature range. This explains why $h/T$ exceeds 1 at high $T$.}
\label{fig:hoft}
\end{figure}

In Fig.~\ref{fig:hoft} we present a sketch of possible temperature evolutions of the Higgs {\it vev}, depending on whether the SNR condition~(\ref{eq:nNcond1}) is met or not and whether the new fermions are sufficiently light compared to the EW scale.
The important variable is in fact the ratio of the Higgs {\it vev} to the temperature, which is a measure of the `strength' of EW symmetry breaking. This turns out to be a key quantity when considering baryogenesis, because the crucial criterium for freezing in the baryon asymmetry is $h/T\gtrsim 1$. When this condition is satisfied, 
sphalerons are not operational and any produced baryon asymmetry during the EW phase transition  cannot be washed out.

In the next section, we refine our discussion and check that our qualitative features   are not altered by higher-order corrections.

\section{A More Refined Analysis}\label{sec:paramspace}

Our analysis of high-temperature  SNR  was so far  limited to the discussion of the leading, one-loop thermal corrections to the Higgs mass. However, the loop expansion in finite-temperature field theory is known for its poor convergence in some cases. In this section, we analyse higher-loop corrections and derive the conditions needed to ensure reliability of the one-loop approximation. 
After deriving the limits of the EFT applicability, we test numerically the allowed parameter space.

\subsection{Finite-Temperature Higher Order Corrections}
\label{sec:fintnlo}

\begin{figure}[t]
\includegraphics[width=11cm]{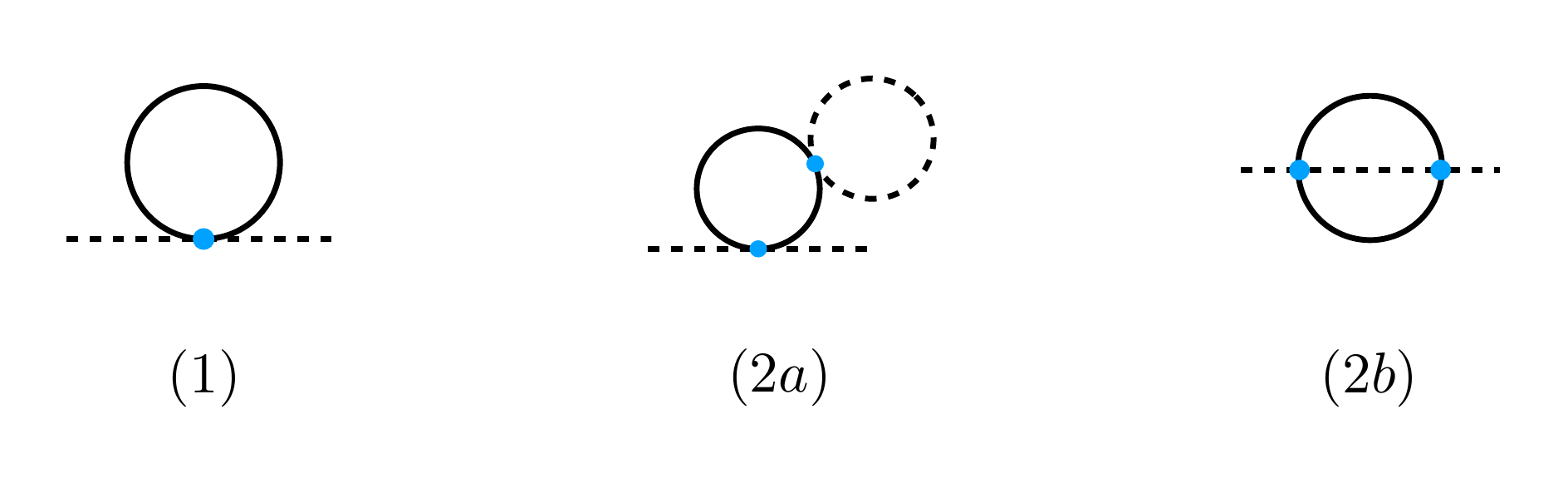}
\caption{The leading one-loop correction to the Higgs mass (1) and two-loop corrections (2a, 2b). Dashed lines correspond to the Higgs boson and solid to $N$.}
\label{fig:nlo}
\end{figure}

First, we remind that the one-loop correction to the Higgs potential (diagram (1) in Fig.~\ref{fig:nlo}) is approximately given by (see  eq.~\ref{eq:deltamh0})
\be\label{eq:alpha1}
\frac{\delta m_h^{\text{(1-loop)} 2}}{T^2}  \sim n \lambda_N \frac{m_N^{(0)}} \Lambda \equiv \alpha.
\ee
and the SNR condition~(\ref{eq:nNcond1}) then reads 
\be\label{eq:alpha1snr}
\alpha   \gtrsim 1. 
\ee
This means that for $n \gg 1$ the SNR condition~(\ref{eq:alpha1snr}) can be fulfilled even for small values of coupling $\lambda_N \propto 1/n$. 
It is exactly this fact that allows to suppress the higher-order loop corrections as we will discuss in the following.

The two-loop corrections to the Higgs mass are given by the diagrams (2a) and (2b) in Fig.~\ref{fig:nlo}. Both can be estimated  as (we suppress the numerical 3D loop factors, see Eq.~(\ref{eq:loopfactor}), see also Appendix~\ref{sec:deltamn1} for the explicit computation of the 1-loop correction to the $N$-mass which is equivalent to the diagram (2a))
\be\label{eq:twoloop}
\frac{\delta m_h^{\text{(2-loop)} 2}}{T^2} \sim n \lambda_N^2 \frac{T^2}{\Lambda^2}.
\ee
First of all, we observe that the relative size of the correction grows with temperature. Such a behaviour is expected for the loop which is induced by a higher dimensional operator, which also shows that our theory unavoidably loses perturbativity at high temperatures. 
Secondly, both corrections are $\propto n\lambda_N^2$, which in the $\lambda_N \propto 1/n$ limit scales as $1/n$. Thereby, these higher-order effects can be effectively suppressed at large $n$ consistently with SNR.   

Let us now discuss more systematically the loop expansion in this theory. A naive guess would be that the maximal possible loop expansion parameter is $n \lambda_N T/\Lambda$. Given that $n$ comes from closed fermionic lines, the dominant sets of higher-order diagrams should be of the daisy type, with multiple fermionic loops attached to an internal scalar line, so that each small coupling $\lambda_N$ is compensated by the multiplicity factor $n$ 
\be\label{eq:n_daisy}
\raisebox{-5mm}{\includegraphics[width=3.cm] {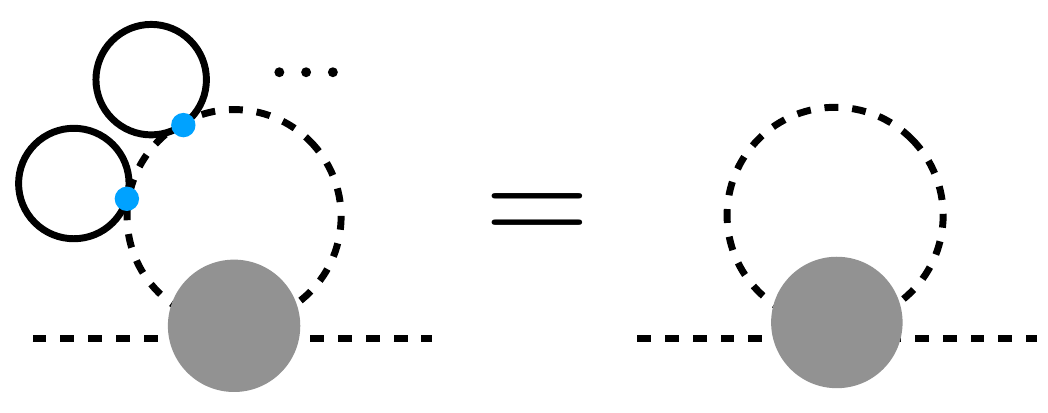}}
\times 
\left\{1+ n \lambda_N \frac{m_N}{\Lambda} \frac{T}{m_h} + \left(n \lambda_N \frac{m_N}{\Lambda}\right)^2 \left(\frac{T}{m_h}\right)^3 + \dots  \right\}
\raisebox{-5mm}{\includegraphics[width=1.7cm] {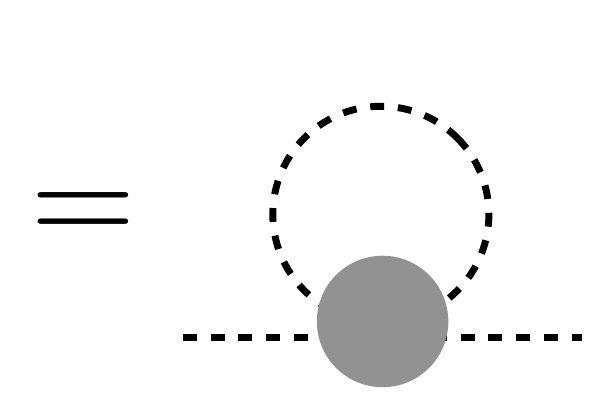}}
\times 
\left\{1+ \sqrt{\alpha} + 1 + \dots  \right\}. 
\ee
Here the gray blob can be anything, but leading contributions would be the Higgs quartic point interaction and the fermionic loop. Such diagrams are IR divergent, hence inverse powers of $m_h$ (see {\it e.g.} Ref.~\cite{Quiros:1994dr} for the power counting in finite temperature QFT). To write the right-hand side, we used the high-temperature expression for the Higgs mass $m_h^2(T)/T^2 \sim \alpha$. We find that the expansion parameter of the series is of order one and is temperature-independent. In principle one may be able to resum such series with some resummation technique. This is however not necessary, as the blob it is attached to has to be suppressed itself, as we now explain. 

All other possible loop series have to scale with a lower power of $n$: any fermionic line, which is not a loop attached to one scalar line, has more powers of couplings for one power of $n$ and is thus more suppressed. The highest loop expansion parameter one can think of is {\it e.g.} that of the series
\be\label{eq:non_daisy}
\raisebox{-4mm}{\includegraphics[width=9.cm] {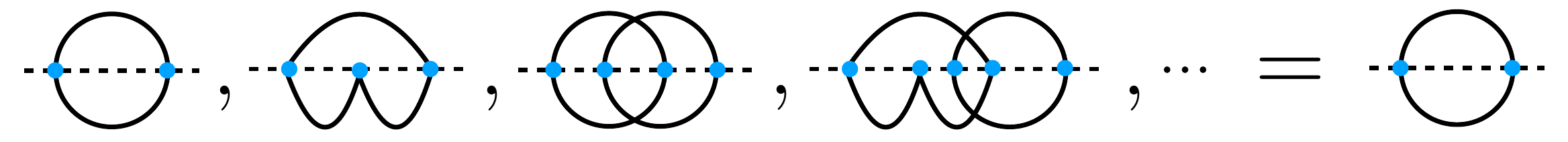}}
\times
\left\{1+ \lambda_N \frac{m_N}{\Lambda} + n \lambda_N^2 \frac{T^2}{\Lambda^2} + n \lambda_N^3 \frac{T^2}{\Lambda^2}  \frac{m_N}{\Lambda} + \dots  \right\},
\ee 
where the alternating $m_N$ factor is required to match the fermion chirality flip induced by the $h^2 N^2$ vertices. 
Such type of series cannot be resummed analytically. We then have to ensure its good convergence, {\it i.e.} require 
\be\label{eq:beta}
\beta = n \lambda_N^2 \frac{T^2}{\Lambda^2} \ll 1.
\ee  
Now, imposing this constraint we see that all the higher loop effects, starting from the two-loop diagrams (\ref{eq:twoloop}) become suppressed. This also applies to the series of fermionic bubbles~(\ref{eq:n_daisy}), as they can only appear on top of some diagrams with $h$ loops, which are by themselves suppressed by $\beta$ or $\lambda_h$. 

A more rigorous way to derive the same conclusions can be for instance by introducing an auxiliary field $\sigma$ mediating the $h^2 N^2$ interaction through $\sigma N^2$ and $\sigma h^2$ vertices, analogously to what is used to analyse large-$n$ $\phi^4$ theories, see {\it e.g.} S. Coleman's lectures~\cite{Coleman:1985rnk}, and what was also applied to scalar SNR in Ref.~\cite{Glioti:2018roy}. It can be shown that in a transformed theory the leading loop corrections correspond to the diagrams with the minimal possible number of $\sigma$ loops. The daisy diagrams~(\ref{eq:n_daisy}), which we identified as a the leading loop series, correspond precisely to the series with no extra $\sigma$ loops, while the subleading series~(\ref{eq:non_daisy}) has an increasing number of $\sigma$ propagators with loop momenta running inside. We do not show this procedure explicitly as it would bring no improvements to the following analysis.

To sum up, we found that the condition $\beta \ll 1$ is necessary for the perturbative expansion to hold. While performing a scan over the model parameters, we will use this condition to define the maximal temperature of the model applicability, 
\be\label{eq:tperturb}
\boxed{T_{\text{max}} = \frac \Lambda {\sqrt{n} \lambda_N}}.
\ee
We will also use a constraint $T<\Lambda/2 \pi$, as the parameter $\Lambda$ by definition sets the scale of new physics which is not captured by our EFT.  The presence of $T_{\text{max}}$ is one of the crucial differences with respect to the scalar SNR scenarios. While the latter are built upon renormalizable interactions, the $\Lambda$-suppressed operator in the fermionic SNR case results in powers of $T/\Lambda$ in the loop corrections, which grow with $T$ and imply a temperature cutoff.

For what concerns the higher-order SM thermal corrections, the leading ones correspond to the corrections to the propagators of the longitudinal SM gauge bosons~\cite{Katz:2014bha}. They are provided in Appendix~\ref{eq:appsm}.

\subsection{Zero-Temperature Corrections}
\label{sec:t0corr}

We complete the description of the loop effects by discussing the zero-temperature quantum corrections.  At one loop, the SM fields modify the Higgs potential by
\be\label{eq:v1lsm}
\delta V_{T=0} = \sum_i (-1)^F \frac{g_i}{64 \pi^2} \left(m_i^4(h) \left( \log \left[ \frac{m_i^2(h)}{m_i^2(v)} \right] - \frac 3 2  \right) + 2 m_i^2(h) m_i^2(v) \right),
\ee
which includes counter-terms ensuring the conditions 
\be\label{eq:renorm}
\partial_h (\delta V_{T=0}(v))=0,\quad
\partial^2_h (\delta V_{T=0}(v))=0.
\ee 
$F=0(1)$ for bosons (fermions), $g_i$ corresponds to the number of degrees of freedom ($g_i=1,4$ for a singlet real scalar and a singlet Dirac fermion respectively). 
Using the same renormalization conditions, we obtain the correction induced by the $N$ loops~\cite{Carena:2004ha}
\be\label{eq:deltav1loop}
\delta V_{T=0}^{(N)} = - \frac{4 n}{64 \pi^2} \left(m_i^4(h) \log \left[\frac{m_i^2(h)}{\mu^2} \right]  + \frac 1 2 c_2 h^2 + \frac 1 4 c_4 h^4 \right),
\ee
with $c_2$, $c_4$ obtained by plugging (\ref{eq:deltav1loop}) into (\ref{eq:renorm})

\bea
c_2 &=& \left\{ \left( - 3 \frac{x x^\prime}{v} + x^{\prime 2} + x x^{\prime \prime} \right) \log \left[\frac{x}{\mu^2} \right] - \frac 3 2 \frac{x x^\prime}{v} + \frac 3 2 x^{\prime 2} + \frac 1 2 x x^{\prime \prime} \right\}, \\
c_4 &=& \frac{1}{2 v^2}\left\{ 2\left( \frac{x x^\prime}{v} - x^{\prime 2} - x x^{\prime \prime} \right) \log \left[\frac{x}{\mu^2} \right] + \frac{x x^\prime}{v} -3  x^{\prime 2} -  x x^{\prime \prime} \right\},
\eea
and $x=m_N^2$. Notice that in this case the non-renormalizable interaction $h^2 N^2$ also generates divergent corrections to the operators $\propto h^6,h^8$. To obtain the expression~(\ref{eq:deltav1loop}) we have fixed the corresponding counter-terms by simply requiring to cancel respective one-loop corrections up to the finite logarithmic terms. As this leaves the potential $\mu$-dependent, we will use $\mu$ as an additional free parameter in the following. 

The loop effects in the presence of new fermions can destabilize the Higgs potential at some $h_{\text{instab}}$ (i.e. the potential would drop below the SM minimum value). We will see in the following that in the parameter space regions which lead to an efficient SNR the instability scale $h_{\text{instab}}$ is always above the cutoff scale $\Lambda$ and also above the $h$ value at which the thermal corrections are minimized. A UV completion of our simplified model at scales above $\Lambda$ then can take care of the instability without interfering with SNR and therefore not affecting the main results of this section.

Now let us discuss the constraints on the applicability of our EFT, caused by the presence of the non-renormalizable interaction $N^2h^2$. The loop corrections would introduce energy-growing corrections to $N^2h^2$, with the expansion parameter
\be
n \frac{\lambda_N^2}{(16 \pi^2)^2} \frac{p^2}{\Lambda^2},
\ee
where $p$ is the typical external momentum. In order for our theory to remain adequate up to the energies $p\sim \Lambda$, we need to impose
\be\label{eq:t0perturb}
\sqrt{n}\frac{\lambda_N}{16 \pi^2} \ll 1\,.
\ee

Importantly, the presence of new physics at the scale $\sim \Lambda$ generically introduces corrections to the Higgs mass of the order
\be
\delta m_h^2 (UV) \sim n \frac{\lambda_N}{16\pi^2} \frac{m_N}{\Lambda} \Lambda^2.
\ee
Further assuming $n \lambda_N m_N/\Lambda \sim 1$, as required for SNR, we get $\delta m_h^2 (UV) \propto \Lambda^2$. The value of $m_h^2/\Lambda^2$ therefore reflects the degree of unnatural fine tuning of the Higgs potential in our model, unless some kind of dynamical Higgs mass adjustment is assumed, {\it e.g.} in the spirit of Ref.~\cite{Graham:2015cka}.

\subsection{Numerical Scan}
\label{subsec:numscan}

\begin{figure}[t]
\hspace{-0.25cm}
\includegraphics[width=0.345\textwidth]{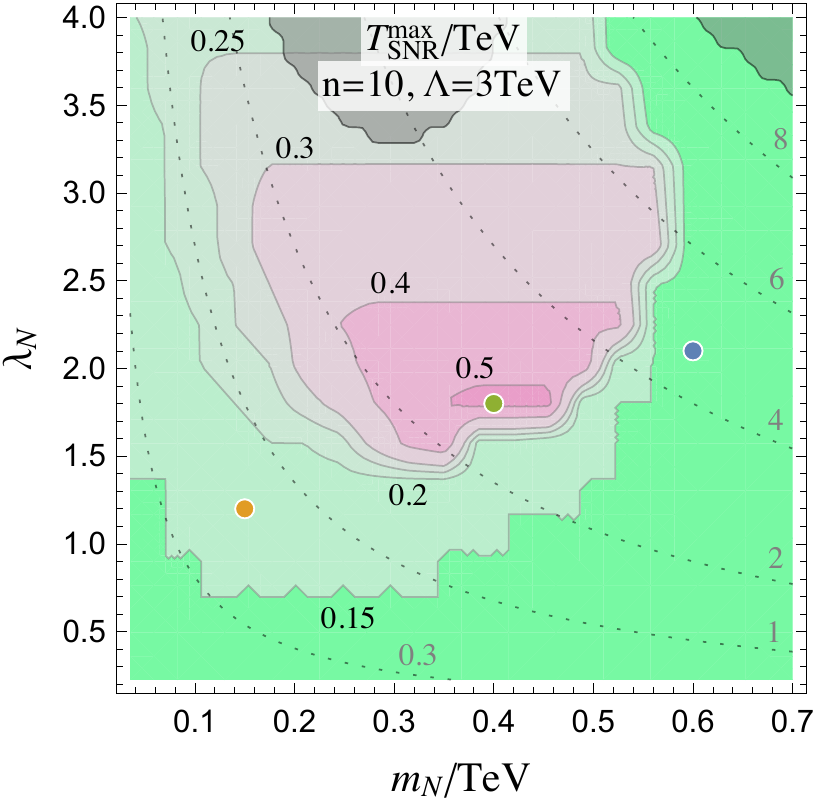}
\hspace{0.25cm}
\includegraphics[width=0.249\textwidth]{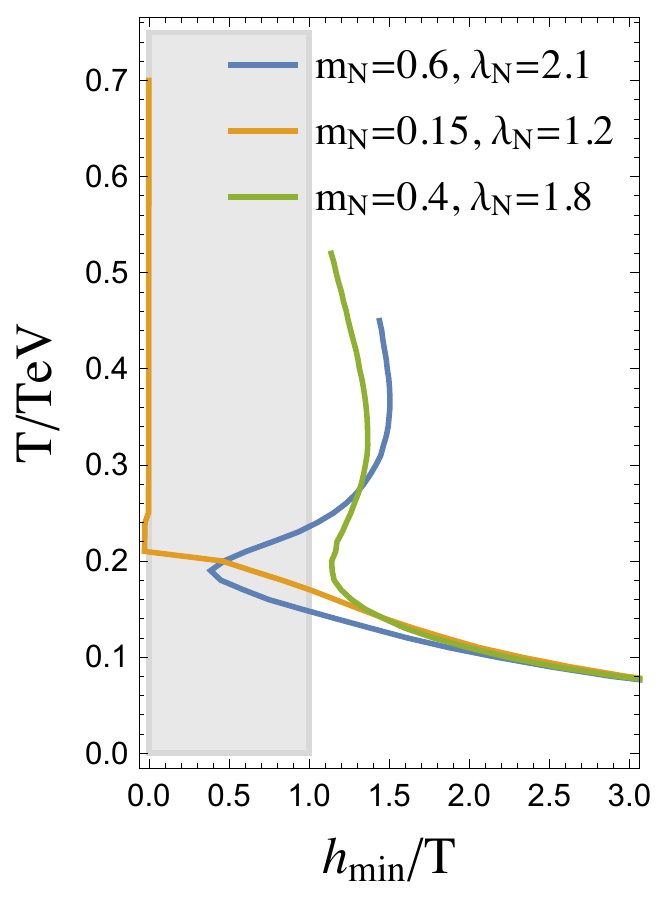}
\hspace{0.25cm}
\includegraphics[width=0.36\textwidth]{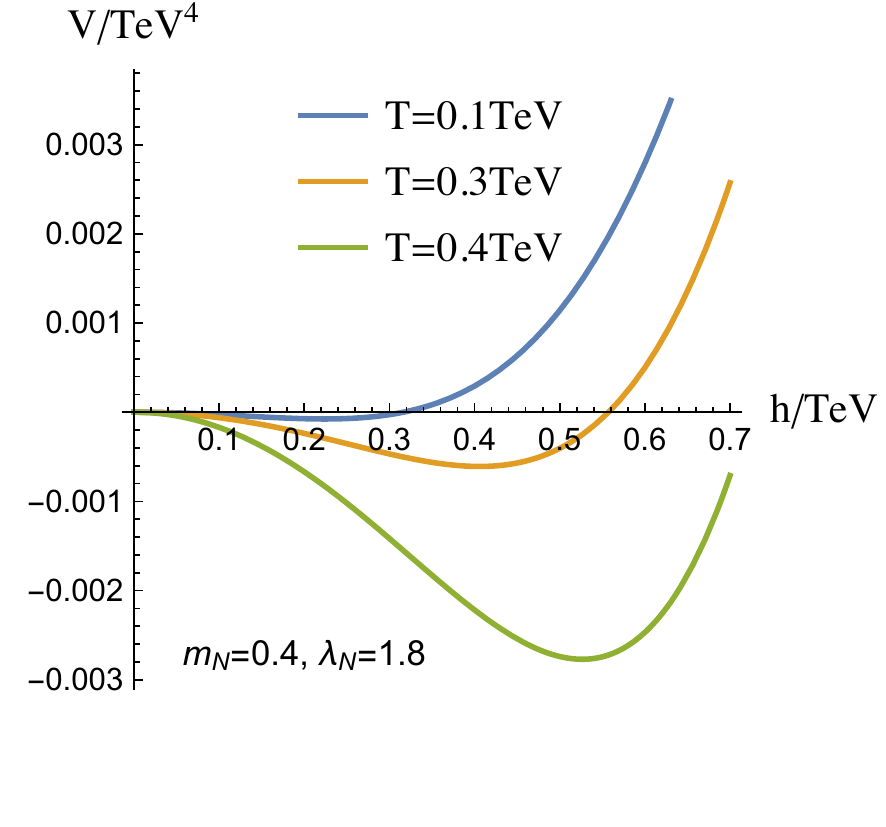}
\caption{{\bf Left}: maximal SNR temperature (colored regions, black labels) for $\Lambda=3$~TeV, $n=10$ and $
\mu=1$~TeV, in terms of the coupling $\lambda_N$ and $m_N$ --  zero-temperature  mass of $N$ at $h=246$~GeV. Grey dotted contours with grey labels show the values of $\alpha = n \lambda_N m_N^{(0)}/\Lambda$. Grey areas feature zero-temperature barriers at $h<m_N^{(0)} \Lambda/\lambda_N$. {\bf Center}: temperature dependence of $h/T$, for three combinations of $m_N$ and $\lambda_N$ (corresponding to the three colored points on the left plot). The $h/T$ lines are limited by the perturbativity from above. {\bf Right}: Higgs potential at $T=0.1,0.3,0.4$~TeV, for $m_N=0.4$~TeV, $\lambda_N=1.8$.}
\label{fig:comb1}
\end{figure}

We present the results of the numerical computation of the $h(T)$ trajectory in Fig.~\ref{fig:comb1}, for a parameter choice $f=1$~TeV and $n=10$. In the left panel, we show the contour plot of $T_{\text{SNR}}$ -- the highest temperature, starting from which the EW symmetry remains broken with $h/T>1$ down to zero temperature. 
Above the maximal $T_{\text{SNR}}$, either the model becomes non-perturbative and violates the constraint~(\ref{eq:beta}), or $h/T$ is less than 1. 

One of the potentially most interesting applications of the above results is for the first-order electroweak phase transition at temperatures higher than the electroweak scale. While we did not attempt to produce the first-order phase transition with the help of $N$ fermions, this task can be achieved in a number of ways. For a concrete example, we can refer to the Composite Higgs set-up, where the electroweak phase transition happens when the Higgs boson is formed from a new strong-sector  confinement phase transition  at some critical temperature $T_c$. 
If $T_c$ is lower than $T_{\text{SNR}}$, the Higgs will land in a symmetry breaking minimum, and remain in the broken phase all the way to $T=0$ as a result of the $N$-induced thermal corrections. This mechanism may allow to realize the electroweak baryogenesis at the electroweak phase transition even if $T_c$ is much higher than $m_W$.

Let us now give a few comments on the behaviour of $T_{\text{SNR}}$ in Fig.~\ref{fig:comb1}.
We see that $T_{\text{SNR}}$ grows with $m_N$ at low $m_N$, as the negative correction to the Higgs mass is proportional to it, see Eq.~(\ref{eq:deltamh0}). However, after $m_N$ becomes too large, the corresponding thermal corrections become ineffective at $T \sim 100$~GeV. In the latter case one can still have SNR at high temperature, but at lower $T$ it is followed by a restoration phase, or a phase with $h/T<1$ (see blue line in $h(T)$ plot). 
$T_{\text{SNR}}$ also initially grows with $\lambda_N$, however after a certain point the perturbativity requirement~(\ref{eq:tperturb}) starts being a limiting factor and $T_{\text{SNR}}$ drops.

The grey area in the upper central part of the $T_{\text{SNR}}$ plot  in Fig.~\ref{fig:comb1} shows where the one-loop zero-temperature Higgs potential features a barrier at $v< h < h({m_N=0})$, where $h({m_N=0})$ is the Higgs value at which the fermion mass vanishes, as defined by Eq.~(\ref{eq:mN0}). This area only covers the regions of a not very efficient SNR, and therefore is irrelevant for our study. In the rest of parameter space such barriers, and the following instability of the Higgs potential, only appear above $h({m_N=0})$ and therefore the new physics, needed to cure the Higgs instability after the barrier, is not expected to affect the results we present.
The grey area in the upper right corner, also having no overlap with the best SNR region, shows where the zero-temperature Higgs potential~(\ref{eq:deltav1loop}) has a barrier at $h < v_{\text{SM}}$.

\begin{figure}[t]
\includegraphics[width=0.38\textwidth]{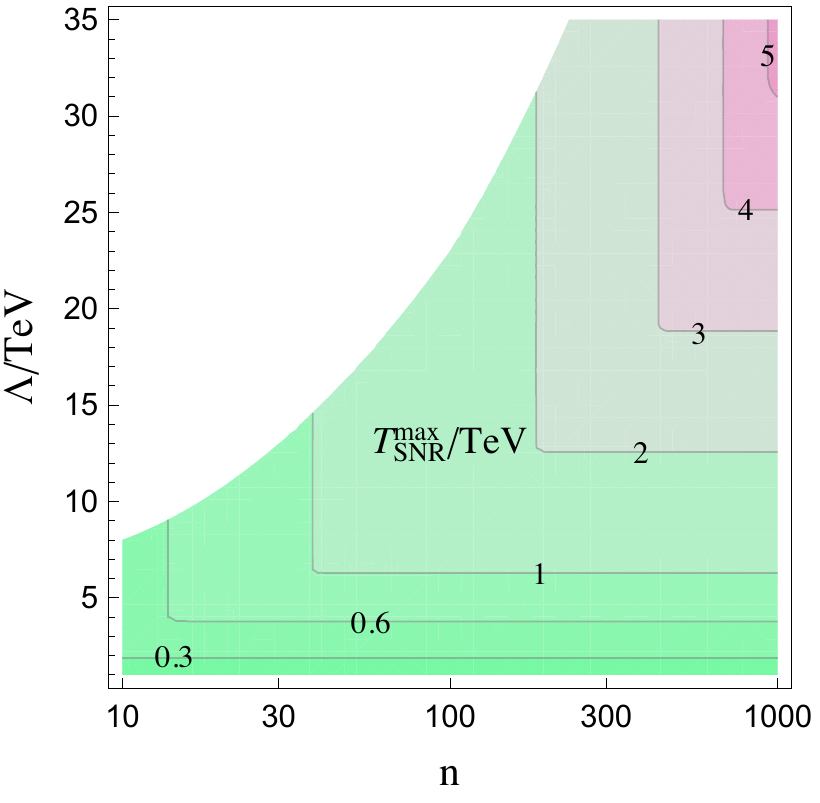}
\caption{Contours of maximal $T$ of SNR (such that $h/T>1$ down to $T=0$), as a function of $\Lambda$ and $n$. In the white area the perturbativity constraint on $\lambda_N$~(\ref{eq:t0perturb}) prevents SNR.}
\label{fig:scan1}
\end{figure}

In Fig.~\ref{fig:scan1} we present the dependence of the maximal $T_{\text{SNR}}$ on $n$ and $\Lambda$, with $\lambda_N$ and $m_N$ chosen to maximize $T_{\text{SNR}}$ in each point. The shape of the contours is mostly defined by two factors. 
First, our theory is not applicable at temperatures above $\Lambda/2\pi$. This defines the horizontal contours in the lower right part of the plot.
Second, the condition to have a negative thermal mass around the origin (see Eq.~(\ref{eq:nNcond1})) together with having $h\gtrsim T$ in the minimum of the thermal correction (defined by $h^2 \simeq m_N^{(0)} \Lambda/\lambda_N$), gives
\be\label{eq:tsnrsm}
\boxed{T_{\text{SNR}} \lesssim \sqrt n m_N^{(0)}\, }.
\ee
The $h\gtrsim T$ condition alone implies $T_{\text{SNR}} \lesssim \sqrt{\Lambda m_N^{(0)}/\lambda_N}$.
The constraint~(\ref{eq:tsnrsm}) defines the vertical contour lines on the plot. Importantly, the perturbativity bound~(\ref{eq:tperturb}) together with the requirement to have a negative thermal mass gives the same expression for the maximal allowed temperature, $T\lesssim \sqrt n m_N$. This means that the non-perturbativity is not a limiting factor for the maximal SNR temperature in our simple model. On the other hand, more involved constructions, such as the one presented in Sec.~\ref{sec:tracking} allowing for a higher $h$ in the minimum, cannot improve on maximal $T_{\text{SNR}}$, as the perturbativity bound remains the same.
A small distortion of the vertical contours at low $n$ and high $\Lambda$ is a consequence of the zero-temperature perturbativity constraint of Eq.~(\ref{eq:t0perturb}).  
In most points $T_{\text{SNR}}$ is maximized at $m_N\simeq 0.3...0.5$~TeV, with the upper limit slightly increasing with $n$. As we will argue in the following, these values are allowed by the current experimental data. 
 
One of the phenomenological constraints on our simplified model comes from the presence of stable SM singlet fermions which interact with the SM only through the $N^2 h^2$ coupling. Analogous models of scalar SNR~\cite{Baldes:2018nel,Glioti:2018roy} were shown to be in tension with the direct dark matter detection experiments. The fixes to this problem include allowing the SNR states to decay either to the SM particles or to a lighter dark matter state. We will leave this topic for future studies as the corresponding modifications, if necessary, can be performed without affecting the SNR. The only remaining experimental signature of the model, which can be discussed in a robust way, is a contribution to the BSM Higgs boson decay rate when $m_N < m_h/2$. Corresponding branching ratio, omitting unimportant numerical and phase-space factors is roughly
\be
\text{BR}_{h \to NN} \sim \frac 1 n \left(n \lambda_N \frac{m_N}{\Lambda} \right)^2 \frac{v_{\text{SM}}^2}{m_h \Gamma_h}, 
\ee
where the parameter combination in brackets is required to be of order a few to provide SNR and $\Gamma_h$ is the full Higgs boson decay width. The requirement $\text{BR}_{h \to NN} < 0.1$ leads to $n\gtrsim10^6$. The only reasonable way to satisfy the experimental data is then to have $m_N > m_h/2$.

\section{UV completions and Generalizations}\label{sec:explmod}

We have argued that the model of Section~\ref{sec:snrwithferm} is the minimal model realising SNR with new fermions, and that the non-renormalizability is a necessary companion of fermionic SNR. 
We will now present a simple argument in favour of this claim, showing that even in more complex constructions the SNR is always related to higher-dimensional operators. We will also get an insight on what the (partially) UV completed theory with SNR should look like, and present two specific examples.

Let us assume we have a theory with some number of new fermions, with the Higgs-dependent masses $m_i$, contributing to the Higgs thermal potential.  
In high-$T$ expansion, their effect on the scalar potential is given by 
\be\label{eq:massarg}
\delta V_f^{T} \simeq  \frac{T^2}{12} \sum_i m_i^2 
= \frac{T^2}{12} \text{Tr}[{\cal M}_{\text{diag}}^\dagger {\cal M}_{\text{diag}}] 
= \frac{T^2}{12} \text{Tr}[{\cal M}^\dagger {\cal M}] 
= \frac{T^2}{12} \sum_{a,b} |{\cal M}_{ab}|^2,
\ee
where ${\cal M}_\text{diag}$ is a fermion mass matrix in the mass eigenstate basis, ${\cal M}$ is the mass matrix in the weak eigenstate basis, and $i,a,b$ enumerate the fermions. In renormalizable SM extensions the matrix elements ${\cal M}_{ab}$ are either Higgs-independent or $\propto h$, so that 
\be
\delta V_f^{T} \propto c_1 + c_2 h^2 \quad \text{with}\quad c_{1,2}\geq0,
\ee
which can only be minimized at $h=0$, thereby leading to high-$T$ symmetry restoration.\footnote{In the case of SNR with new scalars one trivially generates $m_i^2 \propto (\text{const}-h^2)$ from the dimension-four Lagrangian, which allows to produce a maximum of the Higgs potential at $h=0$.}
This conclusion can be overcome if {\it a}) some of the fermion-Higgs interactions are of a dimension higher than four or {\it b}) mass of some of the states entering the mass matrix is much greater than $T$, so that the high-$T$ expansion is not valid. 
We can also conclude that in both cases the temperature at which SNR happens has to be limited from above by {\it a}) the EFT cutoff  or  {\it b}) by the mass of the heaviest fermions. 

We would like to stress that  in the case ({\it b}) the symmetry nonrestoration happens at the temperatures at which some of the fermionic states are very heavy, do not contribute to plasma, and can be integrated out of the theory. After that the remaining low-energy EFT, leading to SNR, is again described by non-renormalizeable interactions, and therefore falls in the category ({\it a}). Whether the theory falls into the category ({\it a}) or ({\it b}), thus depends on the energies up to which the theory is defined. A theory satisfying the condition ({\it a}), however, can also be generated by other types of UV completions, e.g. those with new heavy scalars, or in strongly coupled theories. 

Below we present two specific UV completions to the simplified model, each satisfying one of the two conditions mentioned above.

\subsection{Goldstone Higgs}\label{sec:nghiggs}

In our first example, we will consider the models where the Higgs is a Nambu-Goldstone (NG) boson of some approximate symmetry. In the appropriate parametrization~\cite{Coleman:1969sm,Callan:1969sn}, the Goldstone Higgs appears in the Lagrangian in the form of trigonometric functions which, being expanded, can produce the needed non-renormalizable interactions. 
For definiteness, in our following discussion we will refer to the composite NG Higgs models~\cite{Panico:2015jxa} with the $SO(5)\to SO(4)$ symmetry breaking pattern~\cite{Agashe:2004rs}, while other realizations are also possible.

The Composite Higgs (CH) models can feature both types of the fermionic effects on the thermal Higgs potential discussed in Sections~\ref{sec:modsm} and~\ref{sec:snrwithferm}.
The first effect originates from the top quark Yukawa coupling, which generically takes the form~\cite{Pomarol:2012qf}
\be\label{eq:chyuk}
\lambda_t f \sin^{1+p}(h/f) \cos^r(h/f) \bar q_L t_R,
\ee
where in the following we will take $p=0,r=1$ for definiteness. $f$ is the Higgs ``decay constant'' which is defined by the new strong dynamics. This type of coupling~(\ref{eq:chyuk}) produces the second minimum in the top quark contribution to the thermal potential at $h = (\pi/2) f$. It is important to mention that in the reference PNGB Higgs models the mass of the SM gauge bosons is proportional to $\sin (h/f)$, therefore the second minimum at $h=\pi f/2$ indeed corresponds to the broken electroweak symmetry.   

\begin{figure}[t]
\includegraphics[width=0.24\textwidth]{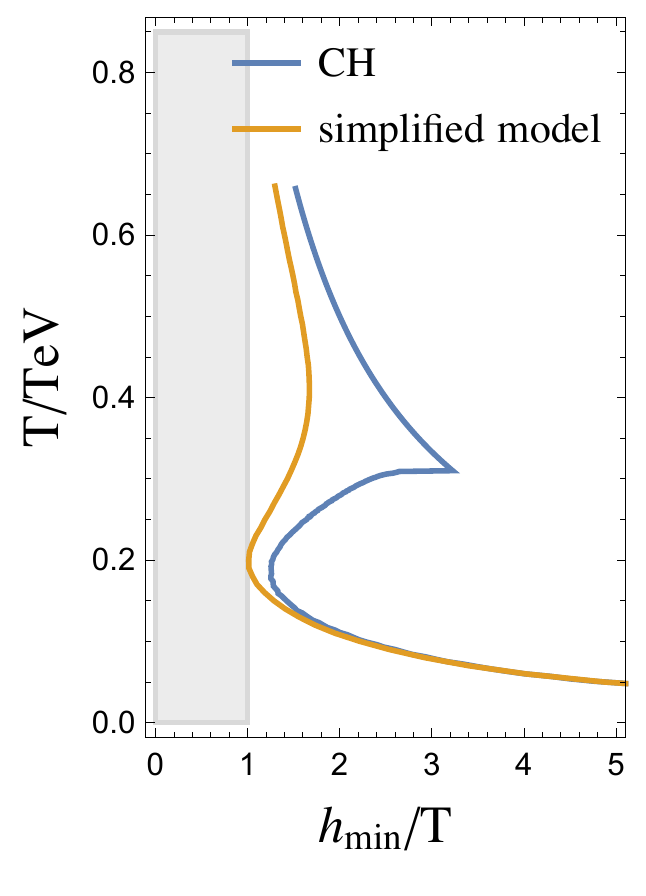}
\includegraphics[width=0.36\textwidth]{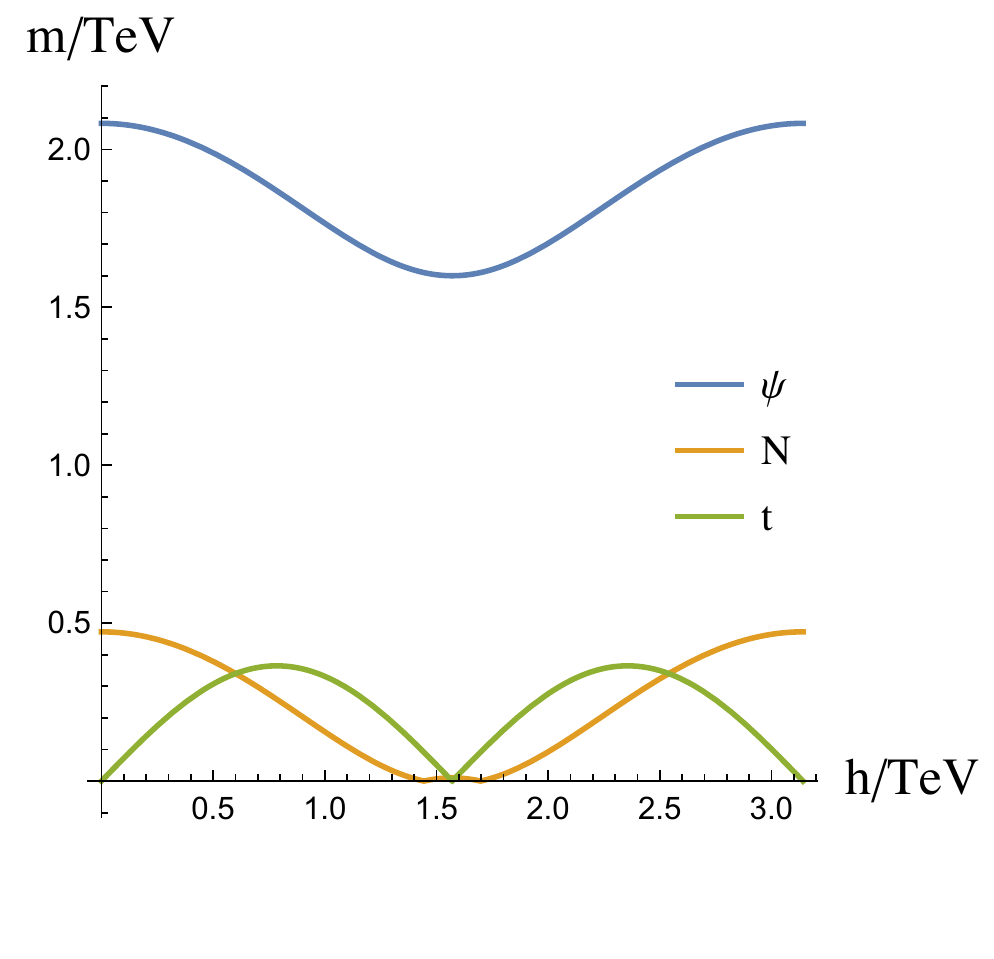}
\includegraphics[width=0.38\textwidth]{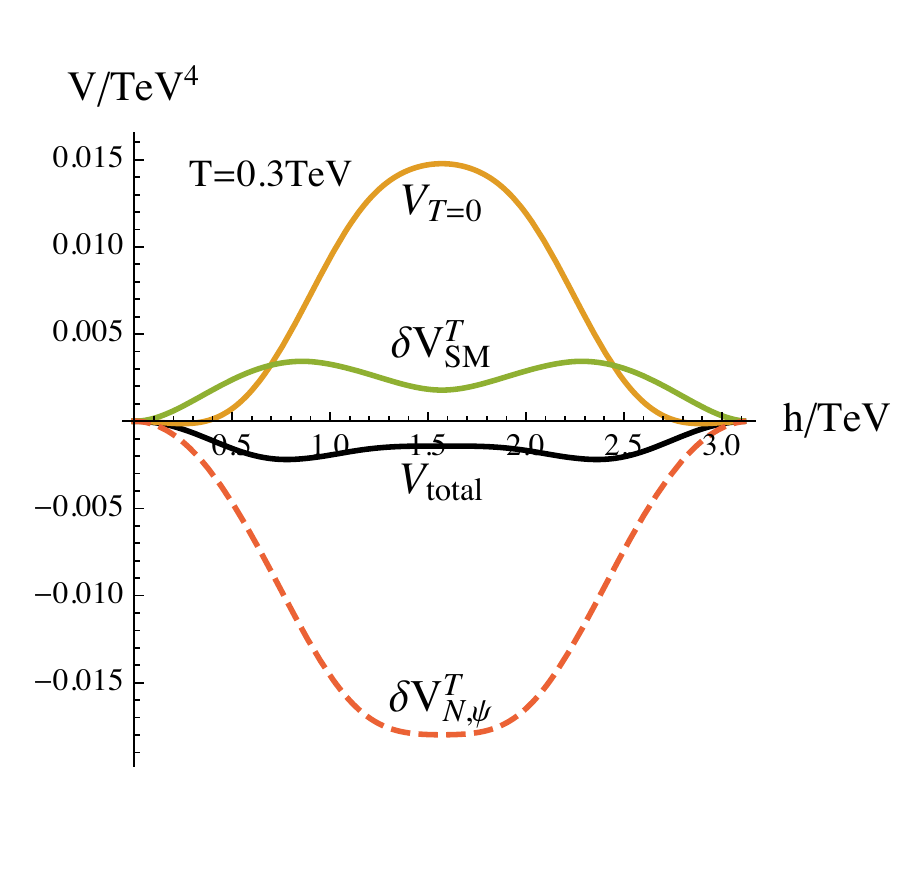}
\caption{{\bf Left}: $h/T$ evolution with a temperature, in the simplified model ($\lambda_N/\Lambda=0.4$~TeV$^{-1}$, $m_N(v)=0.45$~TeV, $\Lambda=2.1$~TeV, $n=15$) and in the respective point of the parameter space of the CH model ($y_L=-y_R=1$, $m_N^0 = 0.1$~TeV, $m_\psi^0 = -1.6$~TeV, $f=1$~TeV). In the CH case we plot $f \sin (h/f)/T$ instead of $h/T$, as EW gauge bosons masses are $\propto f \sin (h/f)$. The $h/T$ lines are limited by the perturbativity from above. {\bf Center}: mass spectrum of the fermions producing the dominant thermal effects. {\bf Right}: overall Higgs potential at $T=0.3$~TeV (black), zero-temperature potential (orange), thermal potential from SM degrees of freedom (green) and thermal potential induced by $N$ and $\psi$ (red dashed).}
\label{fig:ch1}
\end{figure}

The second type of SNR effects can take place if we introduce additional fermionic states into the model. 
In general, even the most minimal CH models do feature new fermionic states -- so called partners of SM fermions, and some of them, such as top partners, can be relatively light~\cite{Matsedonskyi:2012ym,Panico:2012uw}. 
However, their quantum numbers, couplings and masses are constrained by various requirements, such as a need to reproduce the observed SM fermion masses, comply with the electroweak precision measurements~\cite{Grojean:2013qca} and flavour physics constraints~\cite{Barbieri:2012tu,Matsedonskyi:2014iha}, as well as with the bounds coming from the direct searches, which are particularly stringent for the colored partners~\cite{Matsedonskyi:2015dns,Sirunyan:2019sza}.  
For these reasons, we will not try to use SM partners for SNR, but will rather introduce new states. This would give us freedom to choose their multiplicity, quantum numbers and couplings. Let us take an elementary SM singlet Dirac fermion $N$, mixed linearly to its composite $SO(4)$ singlet partner $\psi$. Corresponding mass Lagrangian reads
\be
\boxed{ {\cal L}_{\text{mass}} = f (y_{L} \bar N_{L} \psi_R + y_{R} \bar N_{R}  \psi_L + h.c.) \cos h/f - m_{\psi}^0 \bar \psi \psi - \hat m_N^0 \bar N N\,.}
\ee
where $y_{L},y_{R}$ are dimensionless mixing parameters. 
The determinant of the mass matrix vanishes at 
\be
\cos^2 h/f = \frac{m_\psi^0 \hat m_N^0}{y_L y_R f^2}
\ee
which sets the position of the minimum of the thermal correction. The approximate expressions for the mass eigenvalues (taking $y_L f,y_R f,m_N^0 \ll m_\psi^0$) are
\be\label{eq:chmasses}
m_N \simeq \hat m_N^0 - \frac {y_{L} y_{R} f^2} {m_\psi^0} \cos[h/f]^2\,,\qquad
m_{\psi}\simeq m_{\psi}^0 + \frac{(y_{L}^2+y_{R}^2)f^2}{2m_\psi^0} \cos[h/f]^2 .
\ee
Here and in the following we assume $N$ lighter than $\psi$. 
For completeness, we show the parametrization that we use for the tree-level zero-temperature Higgs potential~\cite{Panico:2012uw} 
\be\label{eq:vCHtree}
V_h = \tilde{\alpha} \sin^2 h/f + \tilde{\beta} \sin^4 h/f\quad \text{with} \quad
\tilde{\alpha}=-2\tilde{\beta} \sin^2(v_{\text{CH}}/f)\,,\;
\tilde{\beta}=\frac{m_h^2 f^2}{8 \sin^2(v_{\text{CH}}/f) \cos^2(v_{\text{CH}}/f)}\,.
\ee
where $v_{\text{CH}}=f \arcsin(v_{\text{SM}}/f)$.
While we call this potential a tree-level, it is supposed to be generated by loops of elementary and composite states. As we are here mainly interested in the general characterisation of SNR, we will not try to model the dynamics responsible for this potential. For the same reason we will also not consider the modifications which may be needed to solve the domain wall problem of PNGB Higgs potentials pointed out in Ref.~\cite{DiLuzio:2019wsw}. 

Using the same renormalization condition~(\ref{eq:renorm}) we derive the one-loop correction induced by the SM states and the new fermions
\be\label{eq:deltav1loopCH}
\delta V_{T=0} = \sum_i (-1)^F \frac{g_i}{64 \pi^2} \left(m_i^4(h) \log \left[\frac{m_i^2(h)}{\mu^2} \right]  + \frac 1 2 c_{2i} (f \sin(h/f))^2 + \frac 1 4 c_{4i} (f \sin(h/f))^4 \right),
\ee
with
\bea
c_{2i} &=& \frac{1}{2f \cos(v_{\text{CH}}/f)^2} \bigg\{ 2\left(f x^{\prime 2} + f x x^{\prime \prime} -2 x x^{\prime} (2\cot(2v_{\text{CH}}/f)+\csc(2v_{\text{CH}}/f)) \right) \log \left[\frac{x}{\mu^2} \right]  \nn \\
&&+3 f x^{\prime 2} + f x x^{\prime \prime} -2 x x^{\prime} (2\cot(2v_{\text{CH}}/f)+\csc(2v_{\text{CH}}/f)) \bigg\}, \\
c_{4i} &=& -\frac{2}{f^3 \cos(2v_{\text{CH}}/f)^2} \bigg\{ 2\left(f x^{\prime 2} + f x x^{\prime \prime} -2 x x^{\prime} \cot(2v_{\text{CH}}/f) \right) \log \left[\frac{x}{\mu^2} \right]  \nn\\
&&+3 f x^{\prime 2} + f x x^{\prime \prime} -2 x x^{\prime} \cot(2v_{\text{CH}}/f)   \bigg\},
\eea
where $x=m_i^2$.
The $h$-dependence of the divergences induced by the SM gauge bosons and fermions, as well as by the new fermions $N$ and $\psi$ is the same as that of the tree-level potential, hence we do not need to include further terms to Eq.~(\ref{eq:vCHtree}) to cancel them. 

In order to make a proper comparison with the simplified model of Section~\ref{sec:snrwithferm}, let us expand the expression~(\ref{eq:chmasses}) in $h/f$. This allows to establish the following relations
\be
m_N^0 \longleftrightarrow \hat m_N^0 - \frac{y_L y_R f^2}{m_\psi^0} \;,\quad 
\frac{\lambda_N}{\Lambda} \longleftrightarrow - \frac{y_L y_R}{m_\psi^0}\,.
\ee
The role of $\Lambda$ is taken by the mass of the heavier state $\simeq m_\psi^0$. Besides $\Lambda$, the scale $f$ also plays role in suppressing high-temperature corrections to the Higgs potential~\cite{Ahriche:2010kh}, we therefore need to impose $T<f$.  

In the left panel of Fig.~\ref{fig:ch1} we show a comparison of the $h(T)$ trajectories for the simplified model and the CH model for one parameters choice. SNR in the CH case is significantly enhanced because of the effect of the top quark. 
We chose $n=15$ for this plot (and not some lower value, e.g. $n=10$, for which the SNR can happen in the simplified model) because it allows for lower $y_{L,R}$ (see SNR condition~(\ref{eq:nNcond1})). At such low $y_{L,R}$ the $T=0$ Higgs potential does not feature any additional unneeded minima.   
We show the mass spectrum of the model in the right panel of Fig.~\ref{fig:ch1}. 

Once embedded into CH setup, the SNR mechanism can have interesting consequences for the electroweak phase transition, allowing to realize it at higher temperature. We can expect that new viable regions of parameter space can be opened for instance in the previously analysed models~\cite{Espinosa:2011eu,Bruggisser:2018mus,Bruggisser:2018mrt,DeCurtis:2019rxl}.

\begin{figure}[h]
\includegraphics[width=0.24\textwidth]{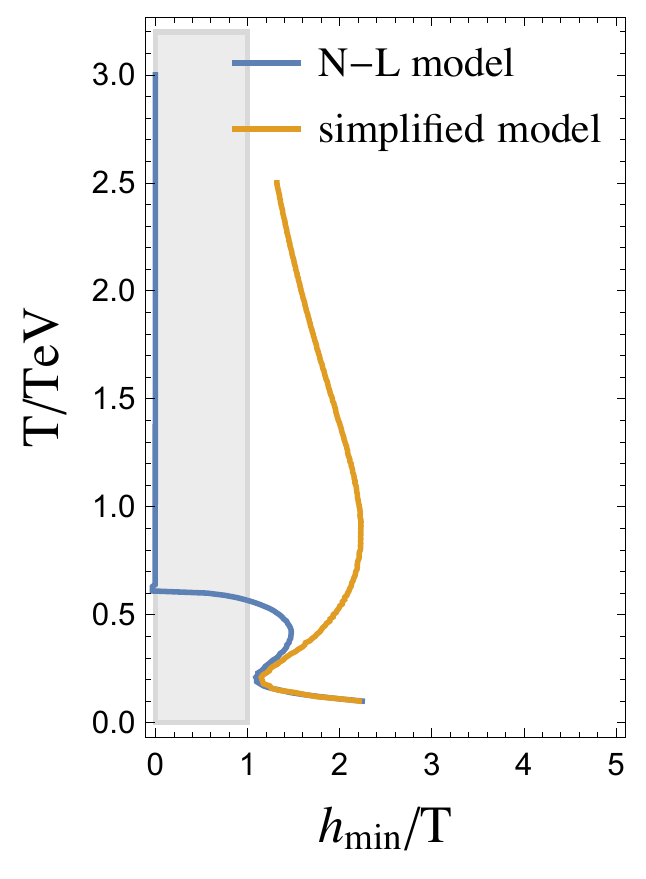}
\hspace{0.1cm}
\includegraphics[width=0.36\textwidth]{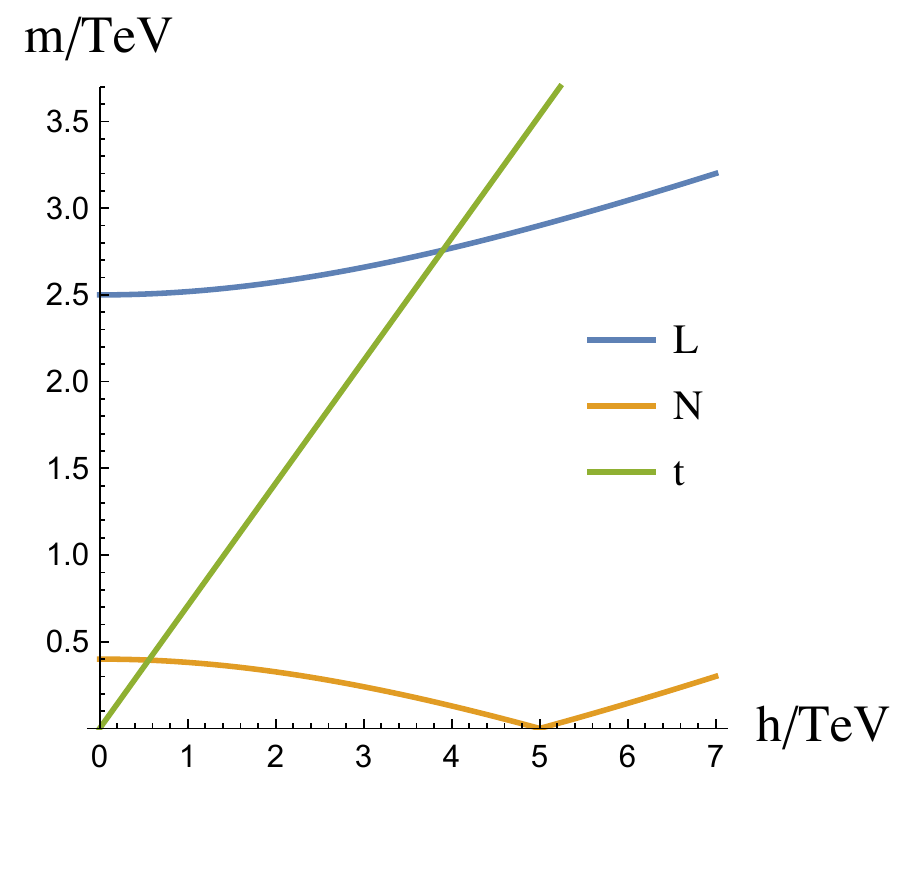}
\hspace{-0.3cm}
\includegraphics[width=0.38\textwidth]{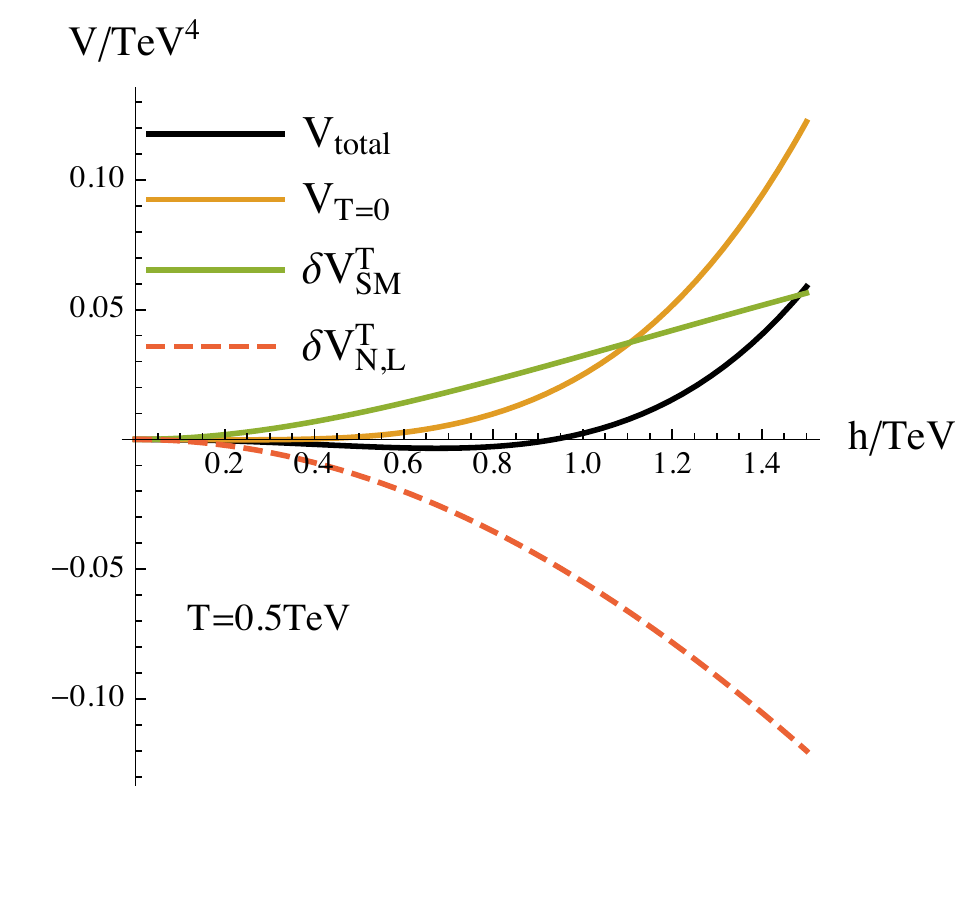}
\caption{{\bf Left}: $h/T$ evolution with a temperature, in the simplified model ($\lambda_N=0.02$, $m_N(v)=0.4$~TeV, $\Lambda=2.5$~TeV, $n=300$) and in the respective point of the parameter space of the singlet-doublet model. {\bf Center}: mass spectrum of the fermions producing the dominant thermal effects. {\bf Right}: overall Higgs potential at $T=0.3$~TeV (black), zero-temperature potential (orange), thermal potential from SM degrees of freedom (green) and thermal potential induced by $N$ and $L$ (red dashed).}
\label{fig:nl1}
\end{figure}

\subsection{Singlet-Doublet Model}
\label{subsec:singletdoublet}

Let us now consider a weakly-coupled renormalizable completion of  the simplified model. We will assume that each of the $n$ singlets $N$ has a heavier Dirac SU(2)$_L$  doublet partner $L=\{L^0,L^-\}$. With the hypercharge difference between the two equal to 1, a tree-level coupling with the Higgs boson is now possible. For concreteness, we fix EW quantum numbers to those of the right-handed neutrino and the left-handed lepton doublet respectively. For simplicity, we also assume a global $U(1)$ symmetry acting on $N$ and $L$, and preventing the coupling of the new fermions to the SM leptons and large neutrino masses. The mass Lagrangian of the new states takes the form
\be
\boxed{{\cal L}_{\text{mass}} = - \hat m_N^0 \bar N N - m_L^0 \bar L L + (y_1 \bar L_L \tilde H N_R + y_2 \bar N_L {\tilde H}^\dagger L_R + h.c.),}
\ee
and the corresponding mass matrix reads
\be
\left[\begin{array}{c}
\bar N_L     \\ 
\bar L^0_{L} \\
\bar L^-_{L} 
\end{array}\right]^T
\left[\begin{array}{c c c}
\hat m_N^0			& -y_2 h/\sqrt 2	& 0 \\ 
-y_1 h/\sqrt 2	& m_L^0		& 0 \\
0			& 0 			& m_L^0 
\end{array}\right]
\left[\begin{array}{c}
N_R     \\ 
L^0_{R} \\
L^-_{R} 
\end{array}\right].
\ee
The electrically-charged state has a Higgs-independent mass and does not affect the thermal Higgs potential. The determinant of the mass matrix of the two remaining states vanishes at 
\be
h^2 = 2 \frac{m_L^0 \hat m_N^0}{y_1 y_2}.
\ee
This defines the point where the thermal Higgs potential can acquire a dip. As we assume that the doublet is heavier than the singlet, the former can be integrated out at low energies, reproducing the simplified model. In the $y_1 h,y_2 h, m_N \ll m_L$ approximation the mass eigenvalues are
\be\label{eq:NLmass}
m_N \simeq \hat m_N^0 - \frac{y_1 y_2 h^2}{2m_L^0} \,,\qquad
m_L \simeq m_L^0 + \frac{(y_1^2+ y_2^2) h^2}{4 m_L^0}.
\ee
The relation between the $N$-$L$ model and the simplified model is given by 
\be\label{eq:nlsmdict}
m_N^0 \longleftrightarrow \hat m_N^0 \;,\quad 
\frac{\lambda_N}{\Lambda} \longleftrightarrow \frac{y_1 y_2}{2m_L^0}\;,\quad 
\Lambda  \longleftrightarrow m_L^0\,.
\ee   

Differently from the model of the previous section, the additional layer of fermions $L^0$ has a mass which grows with $h$, see Eq.~(\ref{eq:NLmass}). Their effect on the thermal potential is then to shift the minimum towards the $h=0$ point. This agrees with the general argument given in the beginning of this section: for the temperature above $N$ and $L$ masses, where the high-$T$ expansion works for both states, the EW symmetry gets restored. SNR can therefore only happen when $L$ mass is sufficiently higher than the temperature. Numerically, the corresponding condition is
\be
T_{\text{SNR}} \lesssim \frac 1 4 m_L.
\ee
However, $m_L$ can not be arbitrarily large. Since $m_L$ is mapped onto $\Lambda$, it suppresses the SNR effect. Concretely, the negative thermal mass condition of Eq.~(\ref{eq:nNcond1}) now reads
\be
n y_1 y_2 \frac{m_N}{m_L} \gtrsim 1.3 \quad \Rightarrow \quad m_L \lesssim 0.8\, n y_1 y_2 m_N\,.
\ee
The combination in the {\it rhs} is further constrained by the perturbativity
\be\label{eq:lnperturb1}
n y_1^2 y_2^2 <1\quad \Rightarrow  \quad n y_1 y_2 < \sqrt n \,.
\ee
This condition can be obtained by expressing Eq.~(\ref{eq:beta}) using the replacement (\ref{eq:nlsmdict}) and accounting for the fact that the $L$ propagators cancel the temperature growth, so that the perturbativity condition is now temperature-independent.

Combining these equations, we finally find
\be
T_{\text{SNR}} \lesssim \frac 1 5 \sqrt n m_N.
\ee
which is smaller by a factor of 5 than the estimate for the simplified model~(\ref{eq:tsnrsm}). This implies a significant increase in the number of new fermions needed for SNR compared to the PNGB model of the previous section.
In Fig.~(\ref{fig:nl1}) we present a numerical comparison of the simplified model and the $N$-$L$ model for one choice of parameters. 
Given that the renormalizable $N$-$L$ model remains a viable description at energies above $m_L$, we do not have to impose the $T<\Lambda=m_L$ restriction anymore. However, a large number of $L$ fermions charged under the weak interactions and a hypercharge may induce a Landau pole at energies not too far from $m_L$. In particular, for the benchmark model used for the plots in Fig.~\ref{fig:nl1}, the weak coupling Landau pole appears around 6~TeV
following from $\Lambda_{\text{L.p.}}\sim m_L \exp{[(4 \pi/ \alpha_2(m_L)) (3/4n) ]}$, where we only show the contribution of the new fermions to the running which is sufficient given their large multiplicity.  
The Landau pole will move to even lower energies at larger $n$.
Furthermore, the loop corrections from new fermions lead to a run-away behaviour of the Higgs potential starting at rather low Higgs field values (we numerically obtain a value $h=6$~TeV for the benchmark model shown in Fig.~(\ref{fig:nl1})), signalling that the $h=v$ minimum is unstable with respect to the quantum tunnelling. This instability can be seen as arising from the running of the quartic term of the Higgs potential induced by the loops of new fermions. The contribution of such loops to the quartic coupling goes as $\propto n y_1^2 y_2^2 = (n y_1 y_2)^2/n \sim 1/n$, where we used the fact that SNR requires $(n y_1 y_2) \sim {\cal O}(1)$. This effect therefore becomes weaker at larger $n$, removing the instability to larger scales.
Both the presence of the instability and a nearby Landau pole mean that this theory requires an appropriate UV-completion already at a rather low scale, discussion of which is however beyond the scope of this paper.

Finally, we would like to point out that similar singlet-doublet extensions of the Standard Model have found many applications in BSM model building, motivated in particular by the gauge hierarchy problem or the baryon asymmetry (for instance in the models of electroweak baryogenesis~\cite{Egana-Ugrinovic:2017jib,Angelescu:2018dkk}, cosmological relaxation of the electroweak scale~\cite{Graham:2015cka} or the weak gravity conjecture-based solution to the gauge hierarchy problem~\cite{Craig:2019fdy}) and therefore SNR may in principle appear as a natural byproduct of these constructions and impact their phenomenology.

\section{Comparison with Scalar-Induced Symmetry Non-Restoration and Temperature Tracking}\label{sec:tracking}

The simplest realization of the electroweak symmetry nonrestoration, already analysed in the recent literature~\cite{Meade:2018saz,Baldes:2018nel,Glioti:2018roy}, is given by the renormalizable models with new scalar fields $\chi_i$ coupled to the Higgs boson. We dedicate this section to the discussion of similarities and differences between these models and our model with new fermions. We parametrize the new scalar fields interactions via the Lagrangian
\be
{\cal L}\, \supset\, -\frac {m_\chi^{(0)2}}{2} \sum_i \chi_i^2 + \frac{\lambda_{\chi h}}{2} \sum_i \chi_i^2 h^2 - \frac{\lambda_{\chi}}{4} \sum_{ij} \chi_i^2 \chi_j^2 \,,
\ee  
where the sums run over $n_\chi$ new scalar fields, for which we assume universal masses and couplings. The first obvious difference is the absence of non-renormalizable interactions, which, on the other hand,  are the reason why the fermionic SNR has an intrinsic temperature cutoff. 
The scalar mass $m_\chi^2 = m_\chi^{(0)2} - \lambda_{\chi h} h^2$ is positive at $h=0$ and decreases as the Higgs {\it vev} grows, thereby adding a negative thermal correction to the Higgs mass 
\be\label{eq:dmhchi}
\delta m_h^2(T) \simeq  - \frac{n_\chi \lambda_{\chi h}}{12}  T^2.
\ee
This correction is active up to the point of the vanishing $m_\chi$, which is determined from
\be\label{eq:minposchi}
m_\chi^2 \equiv m_\chi^{(0)2} - \lambda_{\chi h} h^2 = 0 \quad \longrightarrow \quad h^2 = m_\chi^{(0)2}/\lambda_{\chi h},
\ee
and after this point is crossed the $\chi_i=0$ vacuum becomes unstable. 

For a sufficiently large $n_\chi \lambda_{\chi h}$, the $\chi$ correction to the Higgs mass~(\ref{eq:dmhchi}) can exceed the SM contribution and lead to the stabilization of the Higgs $vev$ away from zero. This is similar to the mechanism at work in the fermionic SNR case. One of the important differences is that the negative cross-quartic $-\lambda_{\chi h}$ may induce an instability of the scalar potential, which imposes a very strict constraint on $n_\chi$. One could in principle try to cure this instability by introducing higher dimensional operators.\footnote{We thank Javi Serra for emphasizing to us this point.} This would obviously introduce a cutoff $\Lambda$, limiting the maximal SNR temperature and the energy of EFT applicability. The cutoff should appear at the field values, at which the potential turns negative, and therefore $\Lambda \propto 1/\lambda_{\chi h}^p$, where the power $p$ depends on the higher dimensional operator that we introduce.  Requiring the negative mass~(\ref{eq:dmhchi}) to exceed the SM contribution, we further derive $\lambda_{\chi h} \propto 1/n_\chi$ and, therefore $\Lambda \propto n_\chi^p$. The cutoff grows with $n_\chi$, similarly to the fermionic SNR that we considered. Derivation of the exact expression for the cutoff for such a model is however beyond the scope of this paper.

Let us now focus on another interesting difference. 
It comes from the effect of the quartic coupling $\lambda_\chi$, which has no counterpart in the fermionic SNR model that we discussed. It induces a one-loop thermal correction to the $\chi$ mass
\be
\delta m_\chi^2(T) \simeq  \frac{(n_\chi+2) \lambda_{\chi}}{12}  T^2.
\ee
So the resulting position of the $m_\chi^2=0$ point actually shifts towards larger $h$ with the temperature:
\be\label{eq:minposchi}
m_\chi^2 \equiv m_\chi^{(0)2} - \lambda_{\chi h} h^2 +\delta m_\chi^2(T) = 0 \quad \longrightarrow \quad h^2 = \frac{m_\chi^{(0)2}}{\lambda_{\chi h} } + \frac{(n_\chi+2) \lambda_{\chi}}{12 \lambda_{\chi h}}  T^2.
\ee
In this way, the $\chi$-induced thermal potential tracks the temperature, and always pushes the Higgs {\it vev} to values $\propto T$, not being limited by any fixed value as in the discussed fermionic model.

We thus find it interesting to discuss the effect of adding four-fermion interaction to our model. It turns out that  such an interaction cannot affect the maximal temperature of a continuous SNR in a significant way. The reason is that even the simple model without the four-fermion interaction is capable of reaching the maximal $T_{\text{SNR}}$ which is allowed by perturbativity, see discussion in Sec.~\ref{subsec:numscan}. The four-fermion interactions can however affect the $h(T)$ trajectory, e.g. by shifting it to higher $h$ values. In order to do so one could add an interaction 
\be
{\cal L} \supset -\frac{c_N} {\Lambda^2} (\bar N_i N_i) (\bar {\tilde N}_j \tilde{N}_j).
\ee
where $\tilde N$ are new singlet fermions and $j=1...\tilde n$. This gives a thermal correction to the $N$ mass (see Appendix~\ref{sec:deltamn2})
\be
\delta m_N(T) = 2 \tilde n \frac{c_N \tilde m_N} {\pi^2 \Lambda^2} T^2 J^F_{F}[\tilde m_N^2/T^2]\;,\qquad 
J^F_{F}[x] = \int_0^{\infty} d k \frac{k^2}{\sqrt{k^2 + x}} \frac{1}{e^{\sqrt{k^2 + x}} +1}
\ee
Using the high-$T$ expansion of this expression, and neglecting the small Higgs-induced thermal correction, we obtain a new fermion mass minimization condition
\be\label{eq:mngap}
m_N[h,T] \simeq m_N^{(0)} - \lambda_N \frac{h^2}{\Lambda} + \frac{\tilde n}{6} \frac{c_N \tilde m_N}{\Lambda^2} T^2 = 0 
\quad \longrightarrow \quad 
h^2 = \frac{m_N^{(0)} \Lambda}{\lambda_N } + \frac{\tilde n}{6} \frac{c_N \tilde m_N}{\lambda_N \Lambda} T^2 \,.
\ee
which is analogous to the scalar SNR case. Therefore maintaining $h/T \geq 1$ at high temperature requires
\be\label{eq:track}
\frac{\tilde n}{6} \frac{c_N \tilde m_N}{\lambda_N \Lambda} \geq 1.
\ee
which is independent of temperature. 
Notice that the same effect could not be achieved with a four-fermion interaction between only one species of fermions $N$. In that case $\delta m_N(T) \propto m_N$ and the new gap equation analogous to~Eq.(\ref{eq:mngap}) would now admit a solution $m_N = 0$ at the same value of $h$ as without four-fermion interactions. Hence the thermal Higgs potential minimum would not shift to higher $h$ as the temperature grows\footnote{It can still shift to lower values for the negative $c_N$. We have checked that the four-fermion operators induced in the two presented UV completions are not dangerous in this respect.}.

In summary, while they are not able to increase the maximal temperature of SNR, the four-fermion interactions satisfying the condition~(\ref{eq:track}) may substantially increase the value of the Higgs {\it vev} at high temperature, and should therefore strengthen the SNR effect.

\section{Conclusions}
\label{sec:conc}

The non-restoration of the electroweak symmetry at high temperature may have a significant impact on the early evolution of our universe. It has been only scarcely studied in the literature so far.
Extending the previous works~\cite{Meade:2018saz,Baldes:2018nel,Glioti:2018roy} on this subject, we have shown that it can take place due to new fermionic degrees of freedom. Among the advantages of this scenario is that it is generally easier to arrange for light fermionic masses, and that new fermions do not alter the Higgs potential at tree level. The intrinsic feature of our scenario --non-renormalizability of the interactions responsible for SNR-- however limits the maximal temperature at which SNR can take place. 

We have analysed the main parametric dependencies of SNR in the framework of a simplified model featuring a single new interaction of the Higgs with a new SM singlet Dirac fermion. The perturbativity of finite-temperature description up to the temperatures ${\cal O}(1)$~TeV leads to the requirement to have at least ${\cal O}(10)$ new SM singlet fermions. The  mass range of the fermions preferred by SNR is $0.4 \pm 0.2$~TeV.  
We have proposed two types of UV completions to this model: the Goldstone (composite) Higgs scenario, and a renormalizable singlet-doublet model. In both cases, the UV completions showed a reasonable agreement with the simplified model.

One of the most interesting implications of  EW  SNR is electroweak baryogenesis, 
a theory of baryogenesis that uses SM baryon-number violation only and relies on a first-order electroweak phase transition. The baryon asymmetry is produced in the symmetric phase in front of expanding Higgs bubbles
though some CP-violating charge transport mechanism and gets frozen as the universe is converted into the broken phase where EW sphalerons are inactive if $h/T\gtrsim 1$. 
While it is relatively easy to trigger a first-order EW phase transition by adding an extra scalar field beyond the SM, the difficulty is that the temperature at which sphalerons freeze-out inside the Higgs bubble is usually around 130 GeV. This imposes EW baryogenesis to take place at relatively low scales, which means that the new CP-violating sources needed for successful baryogenesis are typically in conflict with experimental bounds from electric dipole moments \cite{Andreev:2018ayy}. 
If, on the other hand, one is able to freeze sphalerons much earlier, at higher temperatures, this enables the possibility to realise baryogenesis at higher scales, taking advantages of new CP-violating sources which are less constrained experimentally. 
An interesting potential observable consequence is that the peak frequency 
of the  spectrum of gravitational waves  associated with the first-order EW phase transition is shifted to higher values, which leads to better prospects  for the detectability at  LISA 
\cite{Grojean:2006bp,Caprini:2019egz}.

There are a number of well-motivated extensions of the SM which feature a first-order phase transition at temperatures above the EW scale, up to the TeV scale, involving an extra scalar field. 
Often, EW symmetry does not get broken during such phase transition despite the coupling between the new scalar and the Higgs field, as the temperature of the universe is too high. 
Such conclusion will be changed by involving the mechanism we have discussed in this paper. An EW phase transition may now be naturally induced by the dynamics of the extra scalar field, happening  at temperatures of several hundreds of GeV. EW baryogenesis could then successfully happen. This motivates to revisit such classes of theories.
EW baryogenesis was in particular previously discussed in the context of both types of UV completions that we analysed, see {\it e.g.} Ref.~\cite{Espinosa:2011eu,Chala:2016ykx,Bruggisser:2018mus,Bruggisser:2018mrt,Bian:2019kmg} for EW baryogenesis in composite Higgs models, and ~\cite{Grojean:2004xa,Delaunay:2007wb,Grinstein:2008qi,DeCurtis:2019rxl} for other studies of the EW phase transition in composite Higgs models, and~\cite{Egana-Ugrinovic:2017jib,Angelescu:2018dkk} for EW baryogenesis in minimal singlet-doublet fermionic extensions of the SM. 
In \cite{Bruggisser:2018mus,Bruggisser:2018mrt}, the main prediction for successful baryogenesis in minimal composite Higgs models was a light dilaton, below 700 GeV.  We expect such a bound to be significantly relaxed when adding singlet fermions in the model, largely opening the relevant region of parameter space. These results will be presented in a separate article \cite{Bruggisser:inprep}.

SNR is also generally relevant for models of cold baryogenesis, which are constrained by reheating, such as baryogenesis using strong CP-violation from the QCD axion \cite{Servant:2014bla}.
It may also be interesting to connect this to neutrino mass models with EW scale right-handed neutrino, whose role is played by one of our singlet fermion $N$.
Finally, we stress  that  SNR may impact the phenomenology of recent proposals to address the gauge hierarchy problem such as cosmological relaxation of the electroweak scale~\cite{Graham:2015cka,Choi:2016kke,Banerjee:2018xmn} or weak gravity conjecture-based solutions~\cite{Craig:2019fdy}), both based on
singlet-doublet extensions of the Standard Model similar to the ones we discussed in Section \ref{subsec:singletdoublet}. It would be important to find out whether there are other scenarios of new physics which naturally contain the new states leading to SNR, with one potentially interesting candidate being the Twin Higgs models~\cite{Chacko:2005pe}.

\vspace{1cm}

{\bf Acknowledgements}

We would like to thank Iason Baldes, Javi Serra, Thomas Konstandin and Jos\'e Ramon Espinosa for useful discussions. The work of OM was supported by the IASH fellowship. 
This work is supported by the Deutsche Forschungsgemeinschaft under Germany's Excellence Strategy - EXC 2121 ÒQuantum UniverseÓ - 390833306.

\appendix

\section{Standard Model Thermal Corrections}\label{eq:appsm}

Thermal corrections to the Higgs potential arising in the standard model at one loop level are given by
\bea\label{eq:vhtfullsm}
\delta V_h &=& 
-3\frac{2T^4}{\pi^2} J_f\left[\frac{\lambda_t^2 h^2}{2T^2}\right] 
+ \frac{T^4}{2\pi^2} J_b\left[\frac{-2\mu^2+3 \lambda h^2}{T^2}\right]
+3\frac{T^4}{2\pi^2} J_b\left[\frac{-2\mu^2+\lambda h^2}{T^2}\right] \nn \\
&&
+6 \frac{T^4}{2\pi^2} J_b\left[\frac{g^2 h^2}{4 T^2}\right]
+3 \frac{T^4}{2\pi^2} J_b\left[\frac{(g^2+g^{\prime 2}) h^2}{4T^2}\right],    
\eea
where the third term includes the contribution of three Goldstone modes. The resulting correction to the Higgs mass in the high-$T$ limit is given by~(see Ref~\cite{Katz:2014bha}) 
\be\label{eq:vhexpsm1}
\delta V_h \supset \frac 1 2 h^2 T^2 \left[
\frac{\lambda_t^2}{4} +\frac{\lambda}{2} + \frac{3g^2}{16} + \frac{g^{\prime 2}}{16} 
\right].
\ee
The leading NLO correction to the one-loop result comes from the daisy diagrams. Their main effect can be captured by plugging the thermally corrected masses of the longitudinal EW gauge bosons into the analytic expression for the thermal one-loop Higgs potential~(\ref{eq:vhtfullsm}). The mass corrections are
\be\label{eq:smdaisy}
(\delta M_W^2)_{\text{longit.}} \simeq \frac{11}{6} g^2 T^2, \quad
(\delta M_B^2)_{\text{longit.}} \simeq \frac{11}{6} g^{\prime 2} T^2.
\ee

\section{Temperature Corrections to the Fermion Mass} 
\label{sec:deltamn}

In Sec.~\ref{sec:fintnlo} we performed the power counting of higher-order diagrams, leaving implicit the naive loop suppression factors, which we report here for completeness. Every power of $\lambda_N$ has to be accompanied by a factor 
\be\label{eq:loopfactor}
\int \frac{d \Omega}{(2 \pi)^3}  = \frac{1}{2 \pi^2},
\ee
which we additionally multiply by $4$ for the loops of Dirac fermions $N$.  

We now compute explicitly two types of one-loop corrections to the $N$ mass.

\subsection{$h^2 N^2$ Interaction} 
\label{sec:deltamn1}

Following Ref.~\cite{Quiros:1994dr}, the $N$ fermion self-energy induced by the interaction 
\be
{\cal L} \supset \frac{\lambda_N} {\Lambda} \bar N N h^2 
\ee
at finite temperature is given in the imaginary time formalism by
\be
-i \Sigma_N = \frac{i \lambda_N} {\Lambda} {i}{T} \sum_{n=-\infty}^{\infty} \int \frac{d^3p}{(2\pi)^3} \frac{i}{p^2-m_h^2}
\ee
where $p^{\mu} = \{2 n i \pi T,\vec p\}$ and $p^2 = - 4 n^2 \pi^2 T^{2} - \vec p\,^2$.
We now 
use an identity 
\be
\sum_{n=-\infty}^{\infty} \frac{y}{(2n)^2+y^2} = \frac{\pi}{2}+  \pi \frac{1}{e^{2 \pi y}-1}\,,
\ee
which leads to
\bea\label{eq:sigma1}
-i \Sigma_N &=& 
\frac{i \lambda_N} {\Lambda} 
\int \frac{d^3p}{(2\pi)^3} \left[  \frac{1}{2} \frac{1}{\sqrt{\vec p^2 + m^2}} + \frac{1}{\sqrt{\vec p^2 + m^2}} \frac{1}{e^{\sqrt{\vec p^2 + m^2}/T}-1}  \right].
\eea
The first term reproduces the zero-temperature one-loop correction, which can be completely absorbed into the bare $N$ mass. Using 
\be
\int_{-\infty}^{\infty} \frac{d x}{2 \pi} \frac{1}{-x^2 + \omega^2 - i \epsilon} = \frac{i}{2 \omega},
\ee 
it can be rewritten in the more familiar form
\be
-i \Sigma_N^0 = \frac{ i \lambda_N} {\Lambda} 
\int \frac{d^4p}{(2\pi)^4} \frac{i}{p^2 - m^2+ i \epsilon}\,.
\ee
Remembering that the correction to the fermion mass is given by $\delta M = \Sigma$, we find that the rest of Eq.~(\ref{eq:sigma1}) corresponds to a thermal correction
\be
\delta M_N^T = -\frac{\lambda_N T^2} {{2\pi^2} \Lambda} J^F_{B}[m^2/T^2]\;,\qquad 
J^F_{B}[x] = \int_0^{\infty} d k \frac{k^2}{\sqrt{k^2 + x}} \frac{1}{e^{\sqrt{k^2 + x}}-1}\,.
\ee
For positive arguments, the $J^F_{B}$ function is positive, with a maximal value $J^F_{B}[0]=\pi^2/6$. Also,
\be
J^F_{B}[x] = 2\,\partial_x J_B[x]\,.
\ee

\subsection{$N^4$ Interaction.} 
\label{sec:deltamn2}

Let us now consider the one-loop self-energy correction of the fermion $N$ induced by the four-fermion interactions
\be
{\cal L} \supset -\frac{c_N} {\Lambda^2} (\bar N N) (\bar {\tilde N}_i \tilde N_i),
\ee
with $i = 1...\tilde n$.
The correction is
\be
-i \Sigma_N = \tilde n \frac{i c_N} {\Lambda^2} {i}{T} \sum_{n=-\infty}^{\infty} \int \frac{d^3p}{(2\pi)^3} \text{Tr}[ \slashed p + m_N] \frac{i}{p^2-m_N^2},
\ee
where $p^{\mu} = \{(2 n+1) i \pi T,\vec p\}$ and $p^2 = - (2 n+1)^2 \pi^2 T^{2} - \vec p\,^2$. This expression includes a $(-1)$ factor coming from the fermionic loop. Taking the trace we obtain 
\bea
-i \Sigma_N 
&=& - i 4 \tilde n \frac{c_N m_N} {\Lambda^2} T \sum_{n=-\infty}^{\infty} \int \frac{d^3p}{(2\pi)^3}  \frac{1}{p^2-m_N^2}.
\eea
After applying the equality
\be
\sum_{n=-\infty}^{\infty} \frac{y}{(2n+1)^2 + y^2} = \frac {\pi}{2} - \pi \frac{1}{e^{\pi y}+1},
\ee
the correction reads
\be
-i \Sigma_N 
=  i 4 \tilde n \frac{c_N m_N} {\Lambda^2}  \int \frac{d^3p}{(2\pi)^3} \left[ \frac{1}{2} \frac{1}{\sqrt{\vec p^2 + m^2}} - \frac{1}{\sqrt{\vec p^2 + m^2}} \frac{1}{e^{\sqrt{\vec p^2 + m^2}/T}+1}\right],
\ee
where the first term is the zero-temperature correction
\be
-i \Sigma_N^0 =  i \tilde n \frac{c_N m_N} {\Lambda^2} 
\int \frac{d^4p}{(2\pi)^4} \frac{i \text{Tr}[m]}{p^2 - m^2+ i \epsilon},
\ee
while the second term gives a thermal correction to the mass
\be
\delta M_N^T = 2\tilde n \frac{c_N m_N} {\pi^2 \Lambda^2} T^2 J^F_{F}[m^2/T^2],\qquad 
J^F_{F}[x] = \int_0^{\infty} d k \frac{k^2}{\sqrt{k^2 + x}} \frac{1}{e^{\sqrt{k^2 + x}} +1},
\ee
where $J^F_{F}[0] ={\pi^2}/{12}$ and
\be
J^F_{F}[x] = - 2\,\partial_x J_F[x].
\ee

\end{document}